\documentclass[12pt]{spieman}
\DeclareUnicodeCharacter{22C6}{\ensuremath{\star}}
\DeclareUnicodeCharacter{03BE}{$\xi$}
\DeclareUnicodeCharacter{03B1}{$\alpha$}
\usepackage{phase1_bri}
\usepackage{graphicx} 
\usepackage{subcaption} 
\usepackage{titlesec} 
\usepackage{etoolbox} 
\usepackage{array}    
\usepackage{authblk}  
\usepackage{amsmath}  
\usepackage{hyperref} 
\usepackage[left]{lineno} 
\usepackage{color}
\usepackage{amsmath}
\usepackage{tabularx}
\usepackage{booktabs}
\usepackage{graphics}
\usepackage{epsfig}
\usepackage{enumerate}
\usepackage{pdfpages}
\usepackage{wrapfig}
\usepackage{lipsum}
\usepackage{multirow}
\usepackage{setspace}
\usepackage{color}
\usepackage{comment}
\usepackage{enumitem}
\setlist[enumerate]{itemsep=-0.025in}
\definecolor{mepurple}{RGB}{25, 0, 70}
\definecolor{megreen}{RGB}{25, 120, 50}
\definecolor{red}{rgb}{1.0,0.0,0.0}

\usepackage{transparent}
\usepackage{ragged2e}
\usepackage{titlesec}
\usepackage[font={},justification=justified,format=plain]{caption}
\DeclareCaptionJustification{justified}{\justify}

\newcommand\adeg{\mbox{$^\circ$}}%
\newcommand\amin{\mbox{$^\prime$}}%
\newcommand\asec{\mbox{$^{\prime\prime}$}}%

\usepackage{graphicx}

\usepackage{indentfirst} 

\usepackage{caption}  
\providecommand{\keywords}[1]
{\small\textbf{Keywords---} #1} 

\title{\textbf{Optomechanical Design of the MANTIS SmallSat: An Extreme-, Far- and Near-Ultraviolet Spectrograph for Exoplanet Host Stars}}

\author[1,2,*]{Brian Fleming}
\author[2]{Briana Indahl}
\author[2]{Dmitry Vorobiev}
\author[2]{Kristina Davis}
\author[1,2]{Kevin France}
\author[1,2]{David Wilson}
\author[2]{Nicholas Kruczek}
\author[2]{Timothy Hellickson}
\author[1,2]{Emily Farr}
\author[3]{Marta Civitani}
\author[3]{Davide Sisana}
\author[3]{Kotish Grover}
\author[3]{Laura Proserpio}
\author[3]{Stefano Basso}
\author[3]{Vincenzo Cotroneo}
\author[4]{Fabien Grisé}
\author[4]{Randall McEntaffer}
\author[5]{Oswald Siegmund}
\author[5]{Adrian Martin}

\affil[1]{\small Department of Astrophysical and Planetary Sciences, University of Colorado, UCB 392, Boulder, Colorado, 80309}
\affil[2]{\small Laboratory for Atmospheric and Space Physics, 1234 Innovation Drive, Boulder, Colorado, 80309}
\affil[3]{\small Instituto Nazionale di Astrofisica, Via E. Bianchi, 46, 23807 Merate LC, Italy}
\affil[4]{\small Pennsylvania State University, 201 Old Main, University Park, PA}
\affil[5]{\small Sensor Sciences, 3333 Vincent Road, Pleasant Hill, CA}

\titleformat{\section}{\centering\normalfont\Large\bfseries}{\thesection}{1em}{}

\patchcmd{\abstract}{\quotation}{}{}{}

\begin{document}
\maketitle
\begin{abstract}
The MANTIS (Monitoring Activity of Nearby sTars with uv Imaging and
Spectroscopy) observatory is a compact, multi-instrument small
satellite designed for simultaneous extreme- (EUV; 100 - 560 \AA ), far-
(FUV; 1150 - 1800 \AA ) and near-ultraviolet/visible (NUV/VIS; 2000 - 6400
\AA ) spectroscopy of low-mass stars. The EUV optical system consists of
a first-of-its-kind Hettrick-Bowyer grazing incidence telescope
contributed by the Italian National Institute for Astrophysics (INAF) feeding an advanced e-beam lithographic etched variable line spacing
grating developed at Pennsylvania State University (PSU). The
resulting low-resolution spectrum is imaged on an advanced microchannel plate detector with a potassium iodide (KI) photocathode
for extremely low background noise, resulting in a limiting
sensitivity for MANTIS that exceeds that of the last EUV-sensitive
astrophysics \textcolor{black}{point-source spectrograph, the Deep Survey/Spectrometer (DS/S)} on EUVE. The FUV and NUV/Optical channels
are fed by a compact rectangular telescope that focuses
onto a series of point-source apertures.  The diverging beam is refocused and the FUV band dispersed by a holographic grating, then folded back onto the same detector as the EUV channel by a toroidal fold mirror. The zero-order
light is picked off by a flat NUV grating, with the NUV/Optical spectrum
recorded on an e2v CCD 42-10 detector. The MANTIS spacecraft is a
custom build that leverages the experience derived from prior University of
Colorado - LASP SmallSats for avionics, power, communications, and mechanical
structure. MANTIS is projected to be completed in 2027 with an anticipated 2028 launch as an ESPA-class payload on a rideshare opportunity. 
\end{abstract}

\keywords{astronomy, atmospheres, exoplanets, satellites, optical design, ultraviolet spectroscopy}

{\noindent \footnotesize\textbf{*}Brian Fleming,  \linkable{Brian.Fleming@Colorado.edu} }

\begin{spacing}{2}   

 \section{Introduction}

The search for and characterization of terrestrial exoplanets is one of the three scientific focus areas for astronomy in the 2020s and 2030s, from current efforts to discover atmospheres on rocky planets orbiting red dwarfs with the James Webb Space Telescope (JWST; ~\cite{Redfield24}), to micro-lensing surveys with the soon-to-be-launch Roman Space Telescope~\cite{Terry26}, to the direct detection of Earth-Sun analogs with the upcoming Habitable Worlds Observatory (HWO)~\cite{Feinberg26}.  While those missions aim to understand potentially habitable planets, it is now established that the planetary effective surface temperature does not comprehensively describe the habitable zone. The stellar luminosity alone, which determines the 'liquid water habitable zone', cannot accurately predict and interpret observations of biosignature gases~\cite{Segura05,Tian14}  or evaluate the potential for rocky planets to maintain habitable conditions~\cite{Lammer09,Lammer25}. 

Different ranges of a star’s spectral energy distribution (SED) drive heating and chemistry in different layers of a planet’s atmosphere due
to the wavelength dependence of atomic and molecular photo-absorption
cross sections. Optical and NUV photons combine with the greenhouse effect to heat the surface and lower
levels of Earth-like atmospheres. NUV (1800 – 3200 \AA ), FUV (1000 – 1800
\AA ), and X-ray (05 – 100 \AA ) photons are absorbed in the middle and upper
atmosphere where they can photo-dissociate molecules and stimulate
photochemistry \cite{Harman15}. EUV photons
(10–100 \AA ) are absorbed high in the atmosphere (i.e., in the
thermosphere and exosphere) where they ionize atoms and molecules. Liberated
electrons collisionally heat the surrounding gas, increasing the scale
height and potentially leading to rapid atmospheric escape \cite{Johnstone15}. These EUV
photons are key drivers of atmospheric mass loss as they are absorbed in the lowest density layers of the atmospheres where heating
efficiency is high. 

While great progress has been made
in recent years measuring the FUV and NUV irradiance and variability
due to flares of such stars~\cite{France16,youngblood16,Loyd18,wilson26}, direct observational constraints on the EUV spectra of cool stars remain remarkably rare. The
only previous EUV astronomy mission, EUVE \cite{Bowyer91}, obtained spectra of approximately a dozen cool main sequence stars, including five M dwarfs, over nine years of
operation. The MANTIS (Monitoring Activity of Nearby sTars with uv Imaging and
Spectroscopy) SmallSat is multi-band, multi-instrument spectroscopic
suite designed to characterize the high-energy stellar radiation that
drives atmospheric photochemistry and escape on extrasolar
planets. This is achieved through simultaneous spectroscopy that is
nearly contiguous from 100 -- 6400 \AA , connecting the poorly characterized EUV irradiance of
low-mass stars to the better studied far- and near-UV/visible. With
its broad spectral coverage, MANTIS
calibrates both the quiescent and high energy flare spectral energy
distribution of a selection of M, K, G, and F-type stars
simultaneously.

\begin{figure}[t]
   \begin{center}
   \begin{tabular}{c}
 \includegraphics[width=1.0\textwidth,angle=0,trim={.0in 0.0in 00.0in 0.0in},clip]{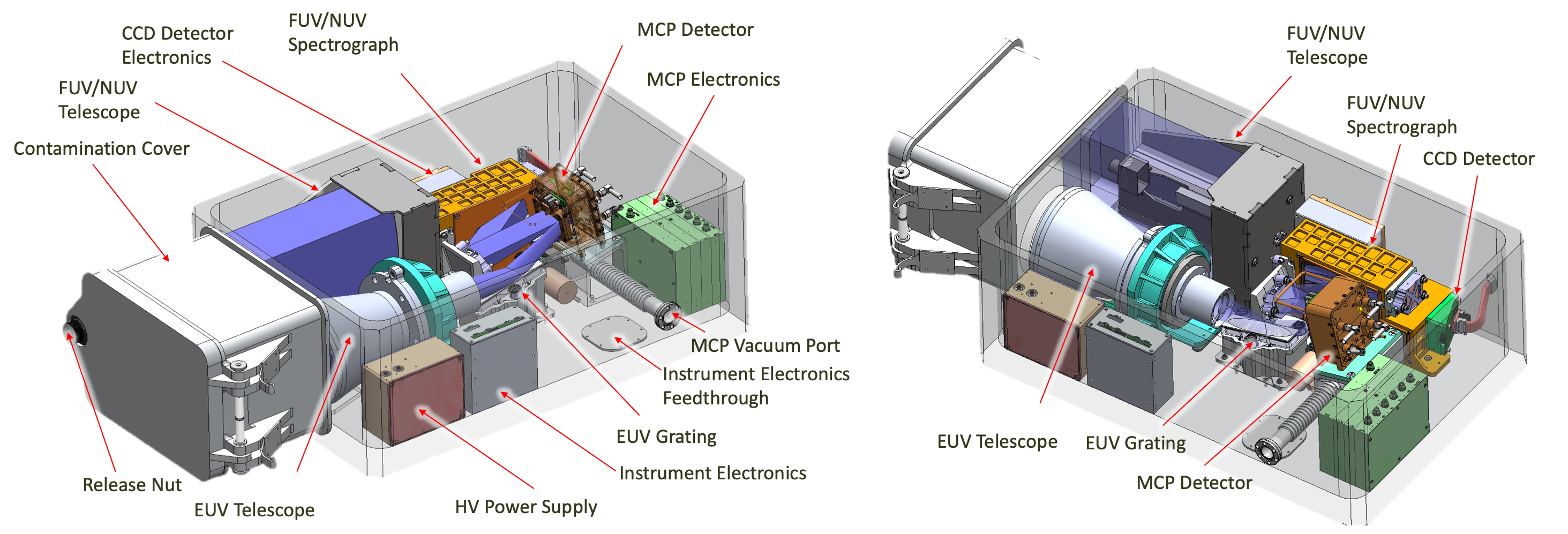} 
   \end{tabular}
   \end{center}
   \caption[Exploded View of the MANTIS Instrument] 
  { \label{fig-layout} 
{\em Exploded views of the MANTIS Instrument. This is all contained within the instrument volume of the spacecraft, with an electrical interface as shown. MANTIS consists of two co-aligned telescopes feeding three spectrographs.}}
\vspace{-0.15in}
\end{figure}

The MANTIS instrument consists of two co-aligned telescopes feeding
three spectral channels, all packaged in an ESPA-deployed SmallSat chassis (Figure~\ref{fig-layout}). The MANTIS instrument and
spacecraft both leverage the heritage of prior University of Colorado -
Laboratory for Atmospheric and Space Physics (CU-LASP) CubeSats,
especially CUTE and SPRITE
\cite{Fleming19,France23,Egan23,Bowen24}, with reflown or slightly
modified detector, optic, and spacecraft hardware.

MANTIS is led by CU-LASP with major contributions from
the Italian Institute for Astrophysics (INAF) and
Pennsylvania State University (PSU). INAF is serving as the EUV telescope lead, providing a first-of-its-kind grazing incidence telescope with electro-formed nickel replicated mirror shells and an optical design optimized for the MANTIS science. PSU is a leader in advanced grating fabrication technologies, with over 10 years of NASA-funded APRA, SAT and Roman Technology Fellowship effort developing silicon diffraction gratings for the x-ray and UV. The PSU team is providing a variable-line-spaced (VLS) grating for the MANTIS EUV-channel fabricated using this technique. Both of these technologies are being flight tested for the first time on MANTIS, and both are also baselined for future large mission proposals (\S\ref{section-conclusion}). 

MANTIS is currently in the third year of development,
having passed through a Comprehensive Design Review in spring
2026. The expected launch readiness date is late-2027 with an expected 2028 launch to a low-Earth
($\gtrsim$ 500 km) sun-synchronous orbit with LTAN between 10:00 -
14:00. Operations will be led out of the CU-LASP smallsat mission
operations group, which features heavy student-operator participation with
professional mentorship, with UHF commanding carried out from the
CU-LASP ground station. X-band science data downlink will be collected
by a third party company which has not yet been identified, though
MANTIS is compatible with several options. The MANTIS program is a
low-budget spaceflight mission funded through the NASA Astrophysics
Research and Analysis (APRA) opportunity.

\begin{figure}

\begin{center}
   \begin{tabular}{c}
 \includegraphics[width=0.85\textwidth, angle=0,trim={0.0in 0in
 0.0in 0.0in},clip]{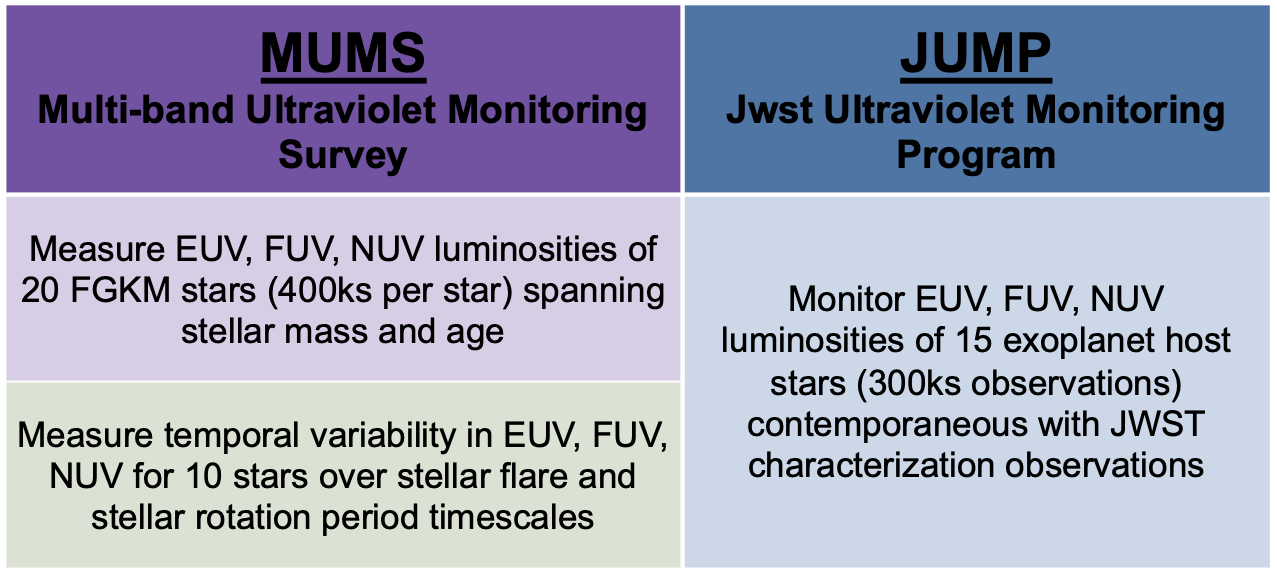} 
\vspace{-0.1in}
\end{tabular}
   \end{center}
   \caption[The MANTIS Science Surveys] 
 { \label{fig-surveys} 
{\em The MANTIS science surveys. The MUMS survey will survey is interleaved between the coordinated JWST observing campaigns that comprise the JUMP survey.}}
\vspace{-0.15in}
  \end{figure}

\vspace{-0.1in}
\subsection{\underline{The MANTIS Observing Programs and Science Concept of Observations}}
\label{section-obs_program}
\vspace{-0.1in}

MANTIS will conduct two surveys (MUMS and JUMP, Figure~\ref{fig-surveys}) to simultaneously characterize the EUV through optical irradiance and time-variability of F, G, K, and M stars. Our concept of operations prioritizes simplicity, building on our team's unique experience conducting science operations with CUTE\cite{France23,Egan23,Patton24}. MANTIS will conduct a ``serial style" observing program, monitoring a single target `continuously' (outside of charging, communications, and keep out avoidance). We refer to these `continuous' observation blocks as campaigns on individual targets, with each campaign totaling an observing time of 100 ks. The 100 ks campaign duration is selected to build EUV signal-to-noise in our fainter targets and to provide the temporal baseline for multi-wavelength flare statistics in our brighter targets.  Factoring in keep-out zones, communication, and target visibility, MANTIS will conduct approximately 3 campaigns per week.

{\it Multi-band Ultraviolet Monitoring Survey (MUMS) }~--~MUMS is the baseline survey that MANTIS will conduct when $JWST$ is not observing an exoplanetary system that is feasible for contemporaneous observations with MANTIS. The MUMS target list is currently being defined by the MANTIS science team and will be selected from a catalog of nearby, bright stars that span a range of stellar masses (spectral types mid-M through mid-F) and ages (ages of a few hundred Myr to ``field age'' (roughly 2~--~8 Gyr; \cite{Schneider19}). The MUMS strategy is designed to measure the UV flux of 20 survey targets at high-S/N in the 12-month primary mission. The targets for the MUMS survey were carefully selected to: 1) have low interstellar hydrogen column densities (log$_{10}$N(HI) $<$ 18.0 for 17-out-of-20 targets based on archival N(HI) measurements with $HST$~\cite{Wood05,youngblood16}, 2) show no evidence of binarity (the white dwarf companion to Procyon will not contribute significant flux at UV wavelengths), and 3) fill out a range of stellar mass and age parameter space, with several ``young'' (age $<$~1 Gyr) and ``old'' (age $>$~1 Gyr) stars in each spectral type bin.

The MUMS observing strategy also samples the variability of high-energy emission on stellar flare and rotational timescales. We execute 4 campaigns per star, spread across the stellar rotation period (all MUMS targets have well-known periods), to sample the rotational modulation of UV flux. The total exposure time per MUMS star is then 4 $\times$~100 ks/campaign = 400 ks per star.  This provides a much longer baseline for cool star studies compared to any previous or proposed study of F, G, and K stars.

{\it JWST Ultraviolet Monitoring Program (JUMP) }~--~JUMP will provide contemporaneous, and in many cases simultaneous, host star observations during $JWST$ transiting planet observations.  MANTIS' dedicated observational platform is designed to provide direct measurements of the stellar conditions that existed prior to and during specific transit datasets obtained with $JWST$.  While we focus here on $JWST$, MANTIS could also support transiting planet observations with $HST$, $TESS$, and even private space missions (e.g., Lazuli, Aperture-1).  Analyzing the exoplanet observing programs $JWST$ executed in Cycle 1, we conclude that MANTIS will be able to sample the full parameter space with both time series F/NUV spectroscopy and, for $\approx$~80\% of targets, EUV flux measurements.  The mission science program and parameters of the joint JWST-MANTIS observing strategy will be outlined in a pre-launch science mission overview paper.  

The JUMP strategy is designed to conduct one stellar campaign per $JWST$ transit observation.  Assume the 3 transit observations per target configuration that is commonly used by $HST$ and $JWST$ Cycle 1, MANTIS will devote (3 x 100ks) 300 ks to each joint JWST-MANTIS target.  Representative $JWST$ cycle 1 transit visit is between 5 and 10 hours (18 - 36 ks; e.g., $JWST$ GO - 1935, 2062, 2319), so the 100 ks provides scheduling flexibility to match up the MANTIS and $JWST$ schedules and to observe the stars for a significant fraction of the planet's orbital phase prior to the start of the transit observation with $JWST$.  This pre-transit baseline will track the quiescent and flare behavior of the star that may influence the spectrum of the planet~\cite{Lecavelier12, Konings22}.  A MANTIS co-investigator at STScI will set-up an automated notification system for communicating $JWST$'s two-week Science Mission Schedule to the MANTIS student operator team at LASP.

\section{Instrument Overview}\label{section-overview}
\begin{figure}[t]
   \begin{center}
   \begin{tabular}{c}
 \includegraphics[width=1.0\textwidth,angle=0,trim={.0in 0.0in 00.0in 0.0in},clip]{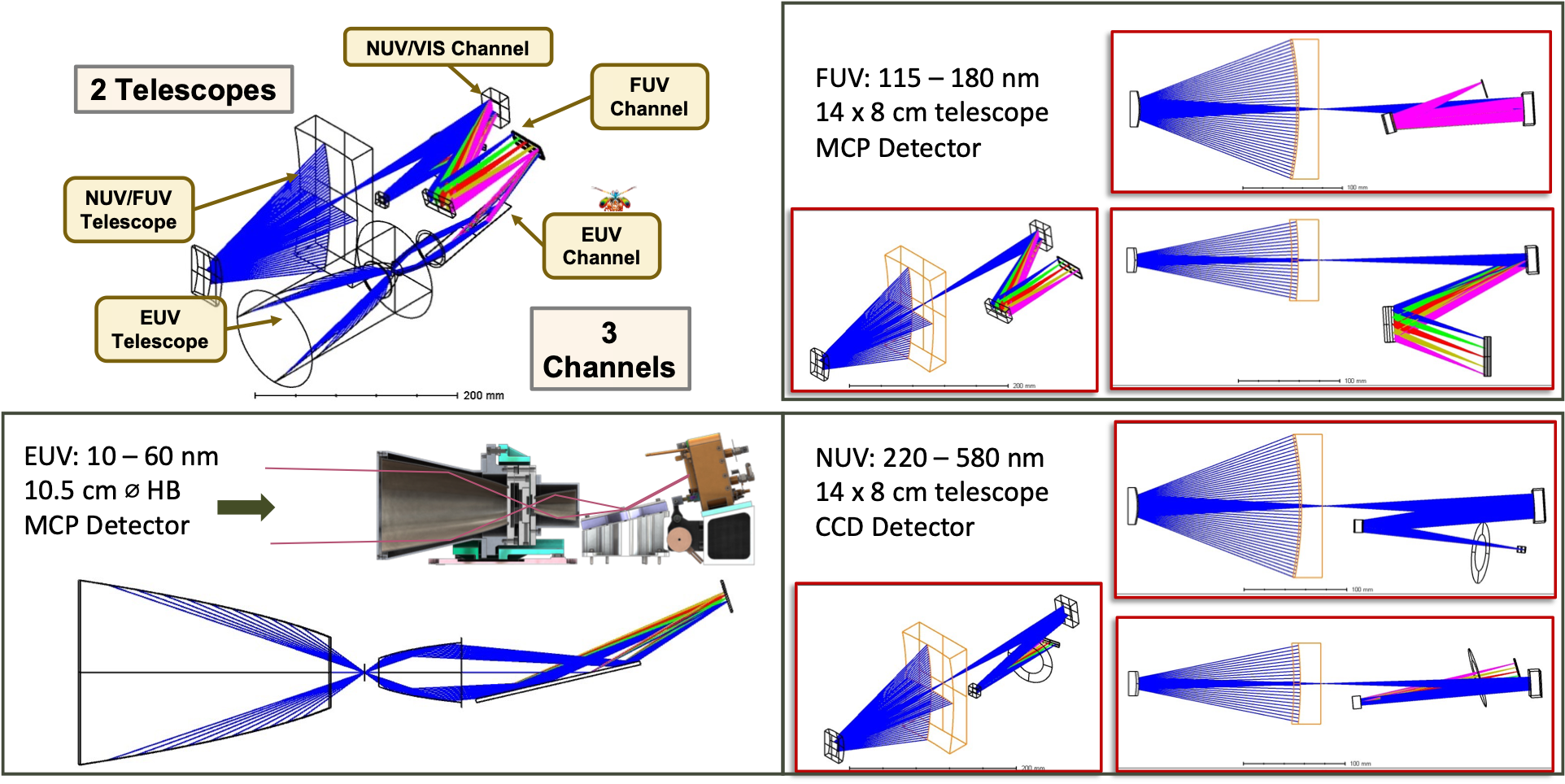} 
   \end{tabular}
   \end{center}
      \vspace{-0.15in}
   \caption[Raytrace view of the MANTIS instrument] 
  { \label{fig-raytrace} 
{\em Raytrace view of the MANTIS instrument. The top left shows the full assembly, with each of the boxes showing only a single channel. The EUV and FUV spectrographs share a common detector, with the spectra occupying different portions of the MCP face. The NUV/O channel uses a CUTE CCD package~\cite{Nell21}. The FUV and NUV/O share a common telescope, with the FUV spectrum dispersed towards the EUV/FUV detector and zero-order dispersed and reflected off of a commercial NUV/O grating.}}
\vspace{-0.15in}
\end{figure}

The MANTIS science instrument suite is divided into three channels,
presented here in order of ascending wavelength: [1] The EUV channel,
[2] The FUV channel, and [3] The NUV/optical (NUV/O) channel. The EUV channel is
mostly independent, consisting of a dedicated telescope and spectrograph
system. The FUV and NUV/O channels share a common telescope
with the two bandpasses split into independent channels at the FUV/NUV/O grating (Figure~\ref{fig-raytrace}). These two channels are
similar in design to prior CU-LASP CubeSats CUTE and SPRITE in both telescope, spectrograph, and detector implementation \cite{France23,Bowen24,Carlson25_2}. The two
telescopes must be boresight co-aligned to within $\pm$1\amin\ for
simultaneous observations (\S\ref{section-coalign}). A small allowance of $\sim$ 5$\%$ is
made for the wings of the point-spread functions (PSFs) to be lost to the
pinholes that define the field-of-views (FOVs).

An overview of the bandpass (both optimized and total),
spectral/angular resolution, and peak effective area is presented in
Table~\ref{tab:performance_specs}, with the architecture of each channel
presented in this section, and a breakdown of the performance metrics
and calculations in \S\ref{section-performance}. 

\begin{table}[!t]
\centering

\begin{tabularx}{\textwidth}{l p{2.9cm} p{2.4cm} X c c c}
\toprule
\textbf{Channel} & \textbf{Aperture} & \textbf{\shortstack{Bandpass\\(Optimized)}} & \textbf{\shortstack{Field of\\View}} & \textbf{\shortstack{Res.\\(Spec.)}} & \textbf{\shortstack{Res.\\(Ang,)}} & \textbf{\shortstack{Peak A$_{eff}$\\($\text{cm}^2$)}} \\
\midrule
EUV & 10.5~cm OD Hettrick-Bowyer & $100 - 560$~\AA\ & 10' Diameter\newline (2' Clear) & 22~\AA & $< 1'$ & 5.1 \\
\addlinespace
FUV & $14 \times 8$~cm Cassegrain & $1150 - 1800$~\AA & 30'' - 60'' Point-source & $\sim 4.0$~\AA & $< 30''$ & 17.1 \\
\addlinespace
NUV & $14 \times 9$~cm Cassegrain & $2000 - 6400$~\AA & 30'' - 60'' Point-source & 7.7~\AA & $< 30''$ & 10 \\
\bottomrule
\end{tabularx}
  \vspace{0.1in}

\caption{{\em A summary of the major performance metrics of each of the MANTIS science channels. The bandpass of each channel is longer than indicated, with the EUV in particular technically covering up to 1100 \AA\ before being vignetted by the edge of the detector, however the resolution and throughput are not optimized for this extended bandpass.}}
\label{tab:performance_specs}
\end{table}

\subsection{The MANTIS EUV Channel}\label{section_EUV}

\subsubsection{The Hettrick-Bowyer Telescope}\label{section-euvtelescope}

\begin{figure}[b]
   \begin{center}
   \begin{tabular}{c}
 \includegraphics[width=0.99\textwidth,angle=0,trim={.0in 0.0in 00.0in 0.00in},clip]{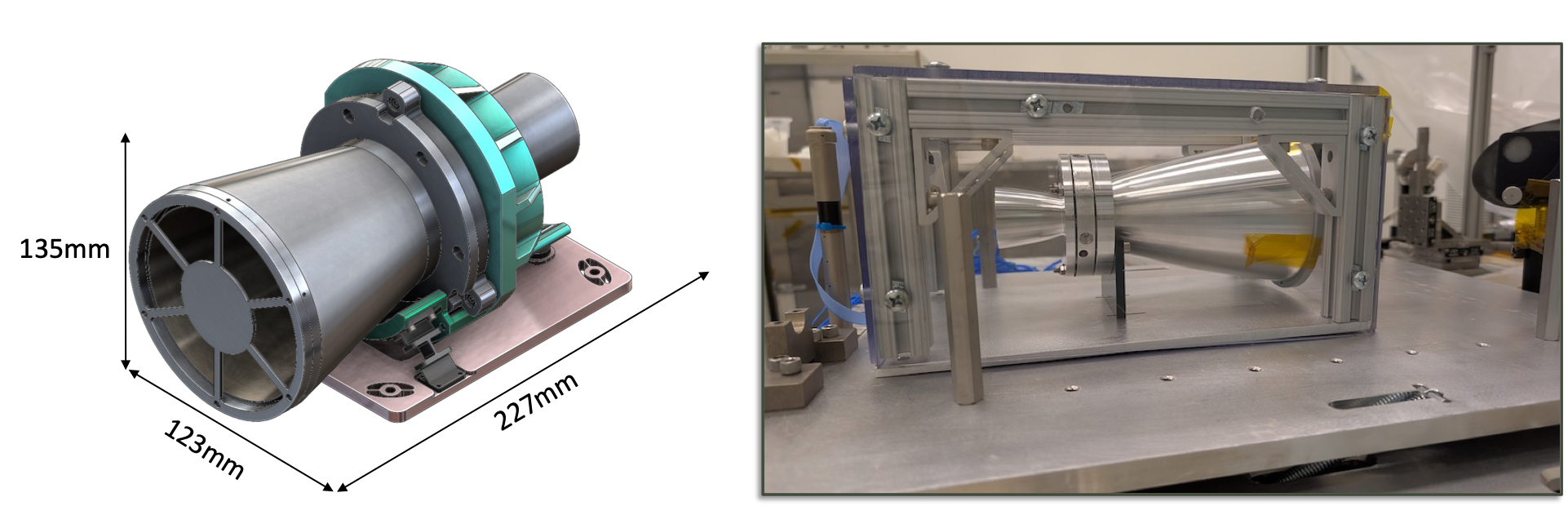} 
   \end{tabular}
 \end{center}
    \vspace{-0.15in}
   \caption[CAD render of the HB telescope]
  { \label{fig-euvscope} 
{\em (Left) CAD render of the EUV Hettrick-Bowyer telescope for MANTIS. The parabolic primary and elliptical secondary mirrors are cantilevered off a central mount, which also contains the prime focus pinhole. (Right) the initial, all aluminum EM from INAF. This unit is a significantly different design from what will be the flight design, but was fabricated to test out alignment procedures and optical test assemblies. This is the first fully integrated HB telescope ever produced.}}
\vspace{-0.15in}
\end{figure}

The MANTIS EUV telescope is the grazing-incidence equivalent of a
Gregorian (known as a ``Hettrick-Bowyer'') \cite{Hettrick84}, with a parabolic primary mirror and elliptical secondary
with a prime focus in-between (Figure~\ref{fig-raytrace}). The Hettrick-Bowyer (HB) was selected for
MANTIS as the two-reflection design has high-throughput with a
compact geometry, while the prime focus enables the placement of an
aperture stop to restrict the FoV to geocoronal \ion{He}{1},
\ion{He}{2}, and \ion{H}{1} emission. While the HB was first theorized
in the era of $EUVE$, one had never been built for scientific
purposes prior to MANTIS. A prototype breadboard was aligned and focused as part of
the recent ESCAPE concept study, from which some of the motivation for
MANTIS was derived \cite{France20,Fleming21}. As of this publication, a MANTIS engineering model (EM) has been integrated, tested, and found to meet the EM-level requirements (Figure~\ref{fig-euvscope}). 

The HB is better optimized for the MANTIS science objectives than the Wolter-II used in the EUVE Deep Survey Spectrometer (DS/S), the previous EUV-sensitive astrophysics spectrograph, as no forward collimator (challenging for a SmallSat volume) is required to restrict the bright geocoronal signal. Even on the night-side, emission lines from the Earth's exosphere are projected to be more than an order-of-magnitude brighter than many of the MANTIS science targets, and the rejection of this emission is one of the primary design drivers of the EUV channel. 

The MANTIS HB telescope is compact
in size, with a 10.5 cm diameter aperture and 380 mm focal length. The
mirrors are fashioned from thin $\sim$ 1 mm thick nickel shells and
mounted cantilevered off a central flange that contains the prime
focus pinhole. Each shell (the primary mirror; PM, and secondary
mirror; SM) are electroformed onto polished aluminum mandrels figured
for a negative of the MANTIS prescription. The telescope and optics
are designed, fabricated and aligned by the MANTIS INAF team at the Osservatorio Astronomico di Brera in Merate, Italy. The INAF team has fabricated, aligned and tested two all-aluminum engineering models (EMs) of the MANTIS telescope to
validate test procedures, with the mandrels for replication in the final polishing stage as of summer 2026. A second EM will also be fabricated using shells replicated from the diamond-turned mandrels prior to the final fine-polishing stage. This EM will use flight-like mounts and be used to validate the optic mounting and alignment plan, as well as for an initial vibration test prior to flight unit fabrication. Despite being only an EM, the point-spread function of the second aluminum telescope already features a half-energy width $<$ 60\asec\ - nearly meeting the flight unit requirement of $<$ 50\asec . 

The prime focus aperture (PFA) is a pinhole of nominally 100 $\mu$m in
diameter, pending the measurement of the PSF of the final MANTIS flight PM. Off-axis light focused by a single grazing incidence
parabola does not behave the same as off a normal-incidence
optic - sources off-center to the PM manifest at the pinhole not as an
off-center focused point, but rather an out-of-focus
``donut''. Therefore, the PFA has an unvignetted transmission for any
source within a diameter of $\sim$ 2\amin , with a 7.4\amin\
diameter total FoV (see \S\ref{section-coalign}) with increasing
aperture losses at larger off-center radii.

\subsubsection{Etched Silicon EUV Variable Line Spaced Grating}\label{section-euvgrating}
\begin{figure}[!t]
   \begin{center}
   \begin{tabular}{c}
 \includegraphics[width=0.99\textwidth,angle=0,trim={.0in 0.0in 00.0in 0.0in},clip]{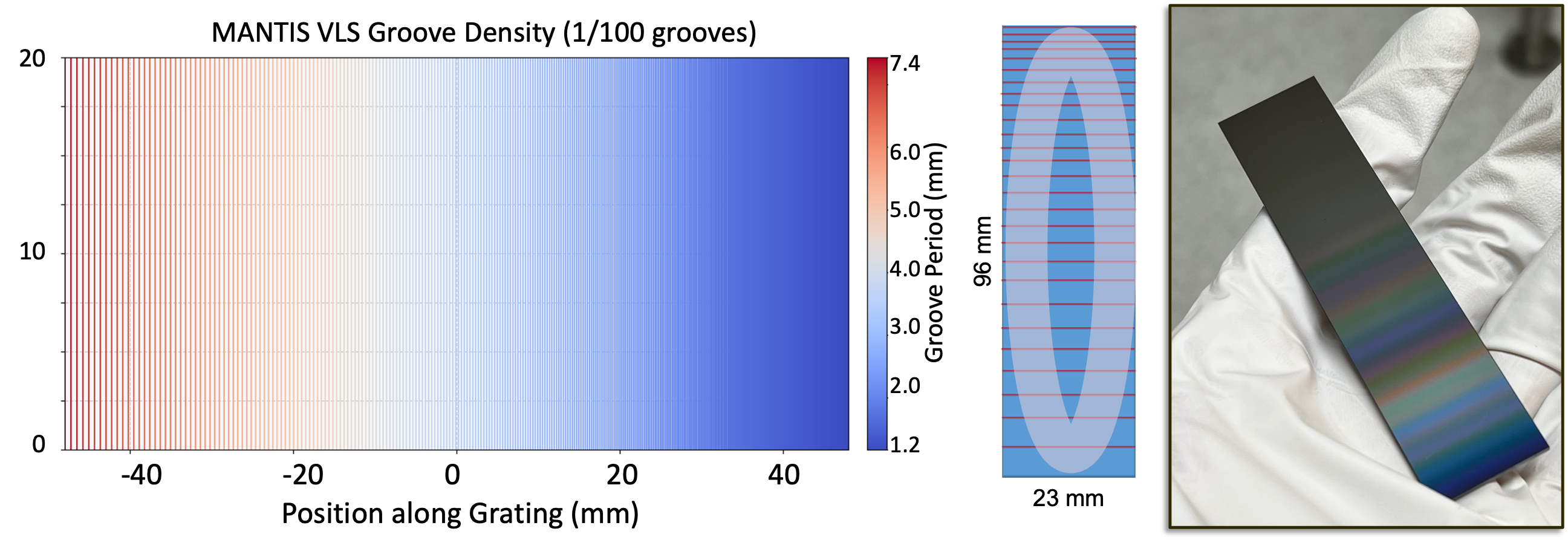} 
   \end{tabular}
 \end{center}
    \vspace{-0.15in}
   \caption[Description of the MANTIS EUV grating]
  { \label{fig-grating} 
{\em (Left) A plot representation of the MANTIS grating groove period as it varies from the side nearest the telescope to the side nearest the detector. The variable line spacing is essential to maintain a constant dispersion at the focal plane in grazing incidence. (Center) a similar cartoon representation of the grating showing how the groove density increases towards the detector. (Right) A photograph of an early MANTIS test grating during the etching phase at PSU. This grating sample was delivered to CU-LASP for testing, albeit with a slightly larger blaze angle (1.4\adeg ) than the design blaze of 0.9\adeg .}}
\vspace{-0.05in}
\end{figure}

The converging beam of the SM is intercepted
by a 96 $\times$ 23 $\times$ 2 mm thick flat, blazed, variable-line-spacing
(VLS) grating (94 $\times$ 21.5 mm ruled area) produced by MANTIS partner PSU. The grating is
fabricated into silicon using electron-beam lithography and etching
procedures initially developed at PSU under several NASA APRA/SAT and RTF
awards \cite{McEntaffer13,Kruczek22,Grise21}. The technology was adapted for MANTIS following over a year of additional development efforts specific to the MANTIS grating needs, which will be highlighted in a future paper (Grise et al., {\em in prep}). The grooves are patterned to
maintain a constant linear dispersion on the detector as the
grating-to-focus distance changes along the grazing incidence
surface, with the groove density spanning 136 - 831 grooves mm$^{-1}$
across the grating length (Figure~\ref{fig-grating}). This creates a very low resolution spectrum at the focal plane, with the
entire EUV spectrum from 100 - 600 \AA\ spanning only 3.8 mm in
length ($\sim$ 131 \AA\ mm$^{-1}$). The low resolution is by-design,
with the intention to maintain a compact signal to reduce the
subtended background on the detector, while shifting the geocoronal
\ion{He}{2} 304 \AA\ and \ion{He}{1} 584 \AA\ emission lines far
enough away from
the 100 - 250 \AA\ and 350 - 500 \AA\ source emission to resolve the stellar
signal. The impact of multiple overlapping spectral orders is discussed briefly in \S\ref{section-euvgrating}, but largely not an issue for MANTIS as spectral information is not as important as the overall photometric irradiance in the EUV. 

The grating is an innovation developed for the MANTIS program and
features an extremely low blaze angle of nominally only 1.0\adeg. Several early prototypes, including 3$\times$ at the full MANTIS size, have successfully produced test gratings with a
1.4\adeg\ blaze angle that have been shown to meet modeled expectations for diffraction efficiency (\S\ref{section-euvgrating}), as well as a single grating with a 0.9\adeg\ blaze angle on a wafer considered too thin for use on MANTIS (250 $\mu$m). Flight substrates with the crystal planes properly oriented for
a $\sim$ 1.0\adeg\ blaze and with the proper thickness have been ordered by PSU with flight gratings expected in the fall of 2026.

\subsubsection{The EUV/FUV Microchannel Plate Detector:} \label{section-detector}
\begin{figure}[b]
   \begin{center}
   \begin{tabular}{c}
 \includegraphics[width=0.99\textwidth,angle=0,trim={.0in 0.0in 00.0in 0.0in},clip]{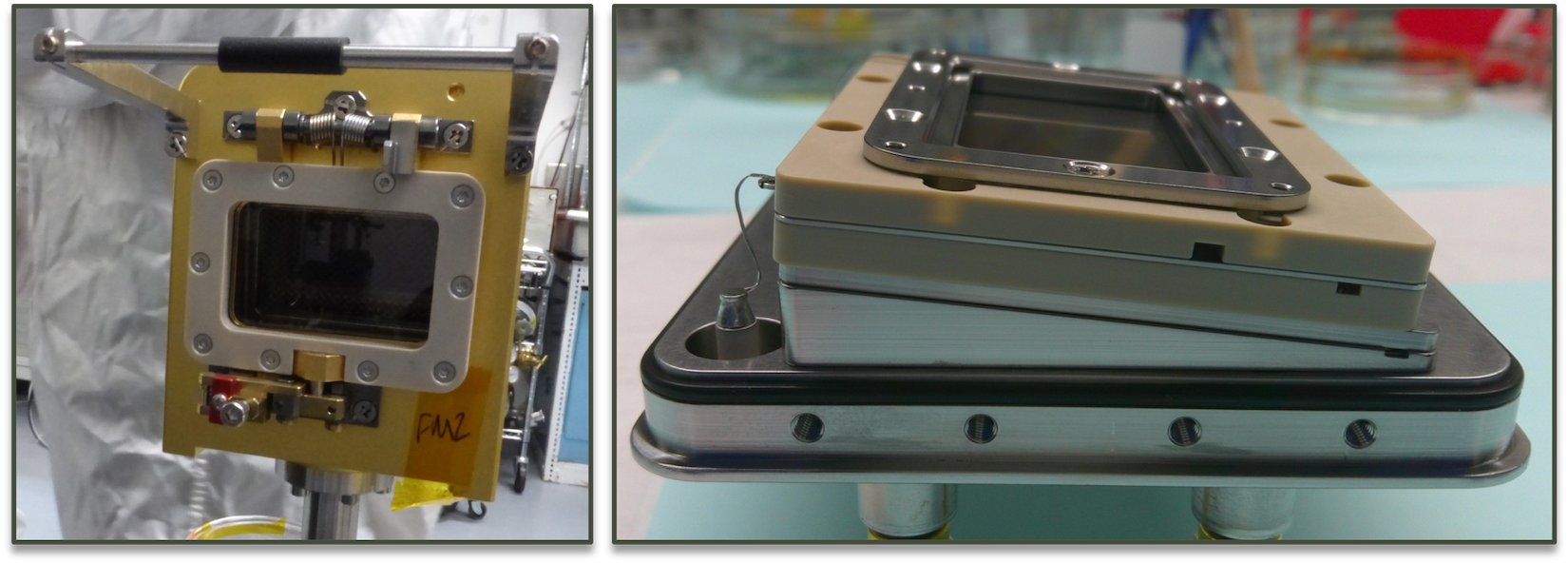} 
   \end{tabular}
 \end{center}
    \vspace{-0.15in}
   \caption[Photograph of the MANTIS flight detector]
  { \label{fig-detector} 
{\em (Left) The MANTIS flight detector in its housing with one-time open door and MgF$_{2}$ window for lab testing. (Right) The MANTIS flight detector out of its housing showing the 6\adeg\ wedge.}}
\vspace{-0.15in}
\end{figure}

The MANTIS EUV and FUV detector is an open face MCP cross delay line (XDL) detector
contained in a hermetic vacuum housing with a manually re-closable door. The
active area of the MCP is 39 $\times$ 17 mm, with the EUV spectral
dispersion direction along the 17 mm axis and the FUV across the 39 mm
axis (Figure~\ref{fig-detector}). The detector has a
solar-blind opaque potassium iodide (KI) photocathode optimized for the
150 – 250 \AA\ EUV bandpass \cite{Ertley26}. This photocathode has
not yet been employed on a flight program, however the CU-LASP FLUID
sounding rocket has received a KI photocathode flight-grade detector with a planned
launch prior to MANTIS \cite{nell2024fluid}. The MANTIS flight and flight-spare detectors have been delivered and tested (\S\ref{section-detector}), with the primary difference between the two being the choice of substrate MCPs. 

The flight detector employs traditional glass substrate MCPs while the flight spare utilizes borosilicate glass, which has flight heritage on large formats. Both use atomic layer deposited activation layers for improved gain stability over the prior generation of lead-glass MCPs, such as those used on $HST$-COS \cite{Green19, Fleming16, Davis18, Siegmund20}. The choice of lead-glass for the flight unit is the result of new advanced plates becoming available that have superior fixed-pattern noise and secondary electron detection efficiency, resulting in enhanced quantum efficiency (\S\ref{section-detector}). A biased, 95$\%$ transmission grid enhances detection efficiency by collecting events that impinge on the front surface web area of the top MCP. 
The detector is mounted on a stainless-steel vacuum flange at a
6\adeg\ angle to help accommodate the tilted EUV spectrum in the
limited volume of the MANTIS chassis (Figure~\ref{fig-detector}). The required detector resolution in the dispersion axis
is $<$ 80 $\mu$m, however the majority of the detector achieves 60 - 65 $\mu$m. MCPs of this type
are capable of higher resolution (e.g. SPRITE)\cite{Bowen24}, however the 80 $\mu$m resolution requirement
is matched to the MANTIS EUV-channel optical performance. The detector readout electronics is similar to the SPRITE and Aspera SmallSat
detector systems, and is derived from those flown on the JUNO and JUICE
Jovian UV instruments \cite{Gladstone17,Bowen24}. 

The detector, electronics, and hermetic housing with manually re-closable door
are provided by Sensor Sciences LLC, the same vendor that provides the
UVS, SPRITE, and Aspera detectors. A sealed housing is essential
to protect the moisture-sensitive KI photocathode and limit molecular
contaminants. This housing contains a MgF$_{2}$ window for
transmission of MANTIS FUV bandpass during testing (Figure~\ref{fig-detector}). The door will
be deployed during vacuum testing, with a custom pin-puller mechanism for in-flight deployment.

\subsection{The MANTIS FUV/NUV/O Channels}\label{section-FUVNUV}

\begin{figure}[!t]
   \begin{center}
   \begin{tabular}{c}
 \includegraphics[width=0.99\textwidth,angle=0,trim={.0in 0.0in 00.0in 0.0in},clip]{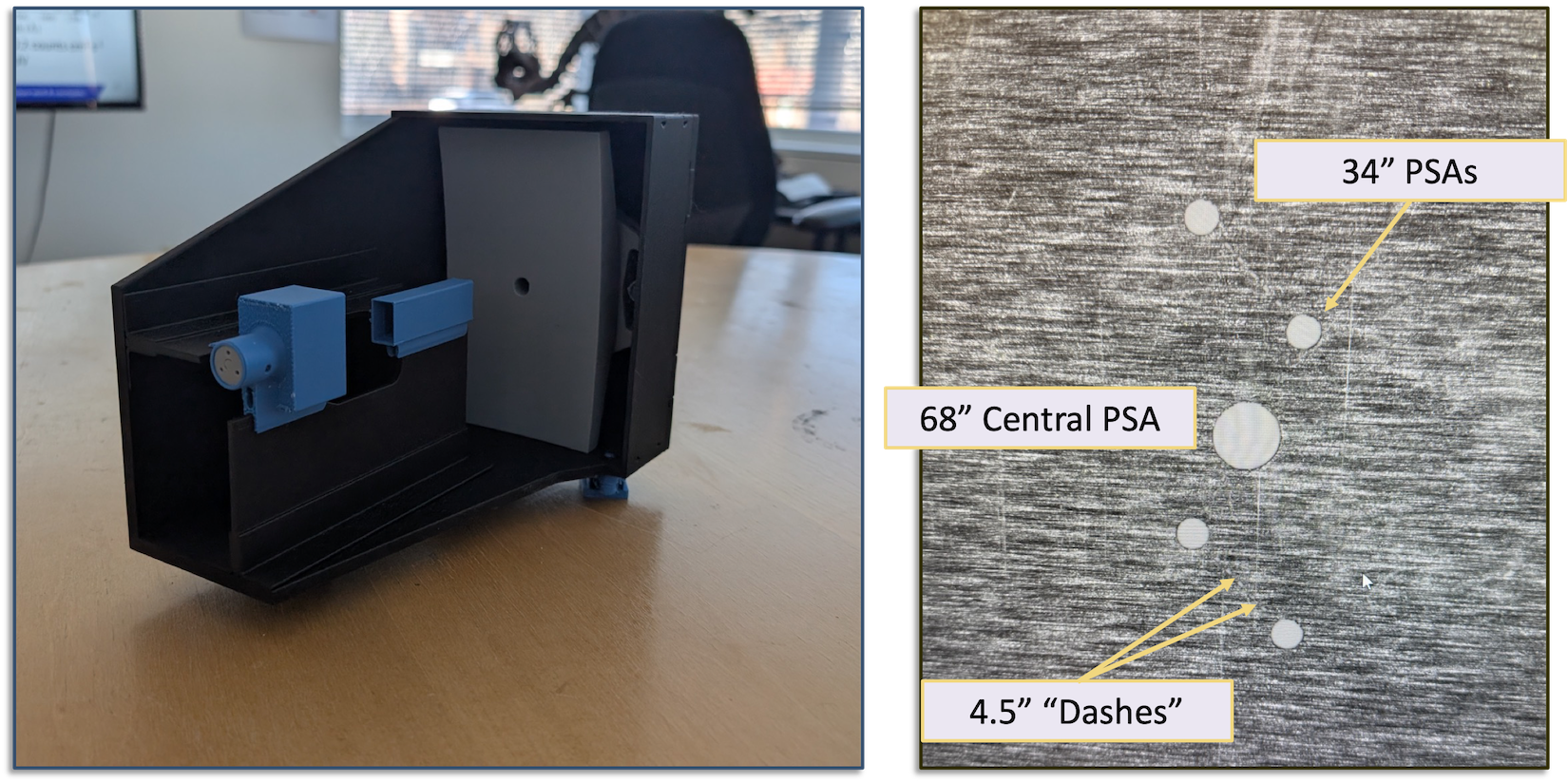} 
   \end{tabular}
 \end{center}
    \vspace{-0.15in}
   \caption[The MANTIS FUV/NUV telescope and slit]
  { \label{fig-teleslit} 
{\em (Left) A 3D-printed model of the MANTIS FUV/NUV/O telescope on a table at CU-LASP. The flight unit is in final assembly. (Right) A magnified photograph of the MANTIS aperture mask. The PSAs measure 200 and 400 $\mu$m in diameter for the smaller and larger PSAs, respectively. As these are only point-source apertures, the mask is fabricated as an electrical PCB stencil into 125 $\mu$m thick stainless steel, providing a cost-effective part with precision sufficient for transmitting point-sources into the MANTIS spectrographs.}}
\vspace{-0.15in}
\end{figure}

The FUV and NUV/O channels are each fed by a 14 $\times$ 8 cm rectangular Cassegrain telescope partially following the design of the CUTE and SPRITE CubeSat telescopes \cite{Fleming17,Egan23,Bowen24}. Unlike those prior telescopes, however, the metering structure of the MANTIS FUV/NUV/O telescope is primarily fashioned from a carbon composite painted with BR-127 black primer for NUV/optical light suppression (Figure~\ref{fig-teleslit}). The telescope comes to a focus at a pinhole assembly just behind the primary mirror, with the beam then entering the FUV/NUV/O spectrograph. The telescope is provided by NuTek Precision Optical Corporation and is coated in MgF$_{2}$ protected aluminum for high reflectivity into the FUV. The telescope optics are polished and the majority of the structural hardware fabricated and painted as of spring 2026. Final alignment of the telescope is progressing with deliveries expected to begin later in 2026, first with EM units and then flight units. 

The light is dispersed by an FUV-optimized grating from Horbia Jobin-Yvon that is a replica of the gratings used in the CU-LASP INFUSE integral-field spectrograph \cite{Haughton25JATIS}. The reuse of this concave aberration-correcting holographic grating represents a significant cost savings for the MANTIS program, however it does drive elements of the optical design as the grating was optimized for the INFUSE payload. The INFUSE/MANTIS FUV grating is 2190 groves mm$^{-1}$ with an 8.2\adeg\ blaze angle for peak efficiency at 1300 \AA . The radius of curvature of the grating is 250.998 mm. The grating is mounted to a custom mount with limited tip/tilt/piston adjustment capability. This is the primary optic for focus and alignment adjustment of the FUV spectrograph. 

{\bf The MANTIS FUV/NUV/O Point-source Apertures:} The MANTIS science is focused on point-source stellar targets, however the aperture mask is designed to provide ample flexibility for co-alignment to the EUV channel, particularly to accommodate launch-related misalignment. Therefore, the MANTIS FUV/NUV/O aperture mask consists of several pinhole/point-source apertures (PSAs) organized in a zig-zag configuration (Figure~\ref{fig-teleslit}). The central PSA is the largest at 68\asec\ diameter, with the other four PSAs all 34\asec\ in diameter. The staggard configuration gives both X and Y direction flexibility to align to the PSA closest to the center of the EUV-channel aperture, ensuring minimal vignetting in either channel. Connecting the PSAs is a 4.5\asec\-wide "slit" made up of dashed apertures which are in place only to help locate the calibration star during early commissioning. 

\subsubsection{The MANTIS FUV Channel}\label{section-fuvchannel}

From the INFUSE/MANTIS grating, the FUV light is dispersed towards the EUV channel and folded back onto the EUV/FUV detector by a custom toroidal fold mirror with R$_{x}$ = 410 mm and R$_{y}$ = 475 mm radii of curvature (Figure~\ref{fig-raytrace}). As with the telescope and grating, the fold mirror is also coated in MgF$_{2}$ protected aluminum. The fold mirror is mounted to a fixed custom optical mount with adjustment by precision shims only as necessary to steer the beam onto the right region of the EUV/FUV detector. All focus adjustment is carried out by the grating, as it is easier to access. 

The FUV spectrum spans from 1120 to 2000 \AA\ over a 39 mm width on the EUV/FUV detector, however the sensitivity of the KI photocathode is less than 1$\%$ at $\lambda$ $>$ 1860 \AA\ and the MgF$_{2}$ optical coatings have reflectance $<$ 15$\%$ for $\lambda$ $\lesssim$ 1140 \AA , resulting in a nearly negligible effective area beyond these limits. 

\subsubsection{The MANTIS NUV/O Channel}\label{section-nuvchannel}
\begin{figure}[!t]
   \begin{center}
   \begin{tabular}{c}
 \includegraphics[width=0.50\textwidth,angle=0,trim={.0in 0.0in 00.0in 0.0in},clip]{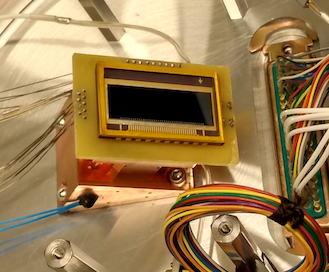} 
   \end{tabular}
 \end{center}
    \vspace{-0.15in}
   \caption[The CUTE CCD as an example for MANTIS]
  { \label{fig-ccd4210} 
{\em Photograph of the CUTE CCD during testing. MANTIS leverages the same CCD and readout electronics for the NUV/O channel detector.}}
\vspace{-0.15in}
\end{figure}

A Spectrum Scientific 300 grooves mm$^{-1}$ flat commercial (COTS) blazed grating intercepts the zero-order reflection off of the INFUSE/MANTIS grating and disperses the longer wavelength NUV/optical light towards a Teledyne/e2v CCD 42-10 detector. This is the same CCD used on the CU-LASP CUTE program (Figure~\ref{fig-ccd4210}) \cite{Nell21}. A Uniblitz F35 35 mm iris shutter lies between the CCD and the NUV/O grating to block light when the CCD is reading out or otherwise not in use. The shutter is set to a normally open mode to maintain the NUV/O channel throughput in the event of a failure. While the NUV/O grating is unpowered, it is mounted with tip/tilt/piston adjustment capability via mechanism shimming to align the NUV/O spectrum on the CCD independent of the FUV grating (which is used to align the FUV spectrum). The focusing is carried out by the curvature of the INFUSE/MANTIS grating, however the NUV/O grating position controls the pathlength to the CCD. 

The resulting spectrum spans the 25.4 mm width of the CCD42-10 with a spectrum spanning 2000 - 6400 \AA . While the NUV/O channel is not sensitive to higher order FUV light, second order NUV will overlap with the optical spectrum starting around 4400 \AA , complicating the optical calibration for the visible portion of the channel for G- and F-type stellar targets. 

\section{The MANTIS Spacecraft}\label{section-spacecraft}
\begin{figure}[!t]
   \begin{center}
   \begin{tabular}{c}
 \includegraphics[width=0.99\textwidth,angle=0,trim={.0in 0.0in 00.0in 0.0in},clip]{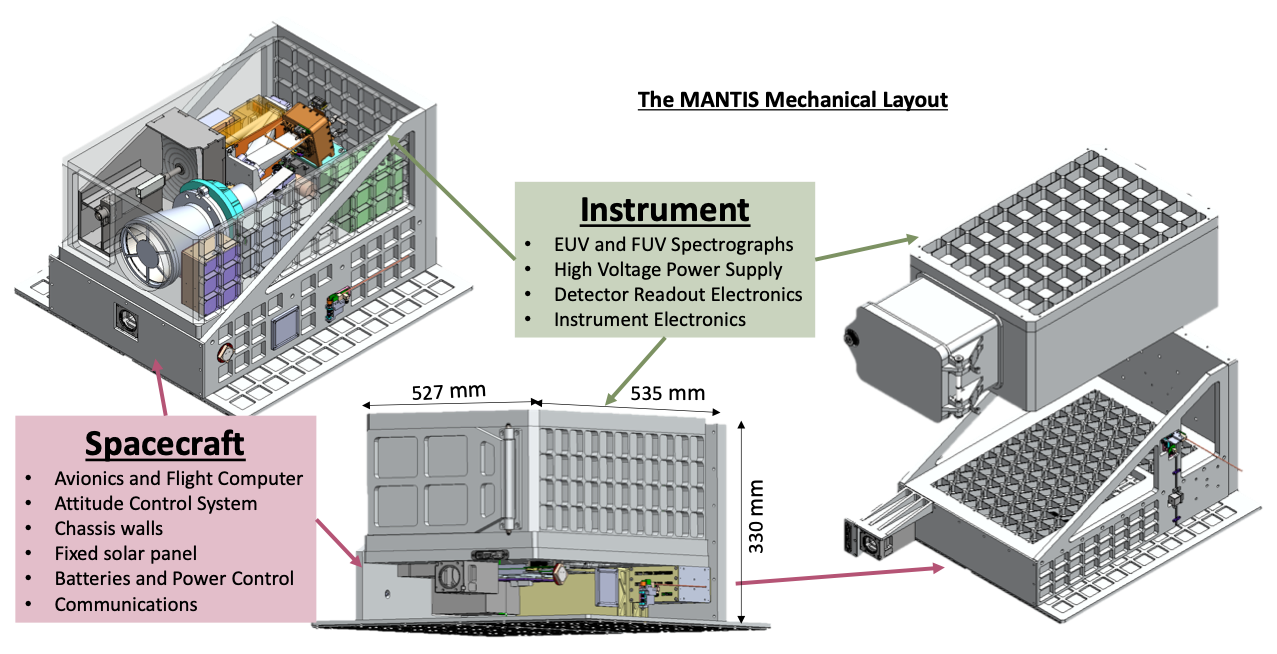} 
   \end{tabular}
 \end{center}
    \vspace{-0.15in}
   \caption[CAD layout of the MANTIS spacecraft]
  { \label{fig-CAD} 
{\em CAD layout of the MANTIS spacecraft showing how the instrument unit and spacecraft unit are distinct, with a single 37-pin micro-D connector between the two. This is a significant reduction in complexity to prior CU-LASP CubeSats where the instrument and spacecraft by necessity are integrated.}}
\vspace{-0.15in}
\end{figure}

The MANTIS spacecraft measures 535 $\times$ 527 $\times$ 330 mm in dimensions and is designed for an 8-inch ring deployed ESPA launch (Figure~\ref{fig-CAD}). The form factor is driven by both cost and simplicity - as the launch costs per unit of mass have drastically fallen in the last decade, there is a significant savings to be had by adding volume to the traditional CubeSat to overcome the complexities of miniaturization. With the MANTIS spacecraft design, the avionics, attitude control, communications, power systems, and other spacecraft components can all be effectively isolated from the science instrument, enabling the two components to be developed in parallel with a simple interface. This is in contrast to previous CU-LASP programs such as SPRITE, where the spacecraft and instrument components were co-located with numerous accommodations necessary to integrate the system. 

\subsection{Spacecraft Avionics}\label{section-avionics}
The MANTIS spacecraft avionics consist of a command and data handling module (C\&DH), Electrical Power System (EPS), generic interface board (GIFB), and an instrument interface board (MIFFY - the MANTIS Instrument Interface Board). The C\&DH, EPS and GIFB are build-to-print, while the MIFFY module is a customized update of the SPRITE Instrument Interface Board \cite{Williams21}, which interfaced with the same Sensor Sciences detector and LASP high voltage power supply module. The primary MIFFY updates are to add more heater control and to interface with the NUV/O-channel CCD, which is an updated version of the CU-LASP CUTE CCD system \cite{Egan23}. The GIFB interfaces with the radios, payload heaters, Blue Canyon Technologies (BCT) XACT-50 ADCS, and other spacecraft hardware. 

The C\&DH contains a Kintex-7 FPGA and a 32 GB 3DPlus rad-tolerant flash memory module. The flight firmware is stored on three redundant TMR configuration flash ICs. The common-code spacecraft flight software (FSW) takes up 0.7 GB of the memory module, leaving 31.3 GB for instrument FSW and on-board data storage. The firmware/software on the instrument electronics is derived from the SPRITE and CUTE programs and includes scripts for filtering unwanted data, trimming or adjusting the active area, and monitoring the detector(s) for count rate, current, and other safety limits. The data output format is in CCSDS packets transmitted to the ground over the X-band radio link. 

\vspace{0.05in}
\noindent{\bf {\underline{MANTIS Power Systems:}}} The MANTIS EPS board contains five regulated voltage supplies, including +28V dedicated to the BCT ADCS and two independent +8V supplies to the UHF and X-band radios. The EPS board also delivers from the solar panels the +14.4 - 16.8V bus voltage that charges the 2$\times$ 93 W-hr LASP-built custom battery modules containing Panasonic NCR18650B battery cells.

The single solar panel is custom built at LASP and mounted to the base of the spacecraft chassis, without any need for deployment once launched. The panel consists of eight nine-series (9s) two-parallel (2p) cell strings. The solar cells are Azur Space triple junction 3G30A cells generating an end-of-life (EOL) power at 80\adeg C of 73 W-hr per panel. Based on the orbit-averaged power of 52.4 W-hr, MANTIS has $\gtrsim$ 20$\%$ power-positive EOL margin over each orbit for up to a 30\adeg\ sun angle of incidence onto a single solar panel. Nominally, the spacecraft will charge every orbit, though MANTIS is capable of observing continuously for several orbits without charging. 

\vspace{0.05in}
\noindent{\bf {\underline{MANTIS Communications System:}}} MANTIS carries two radios - UHF uplink/downlink configured for 450 MHz for housekeeping, command and control, and X-band downlink configured for 8.025 GHz for an estimated 18 Gb/day of data capacity. MANTIS also carries an Iridium radio for limited instantaneous communication. The MANTIS maximum anticipated data rate is less than 4.6 Gb/day. 

\vspace{0.05in}
\noindent{\bf {\underline{Blue Canyon XACT-50 ADCS:}}} MANTIS requires precision pointing control to less than 20" such that the stellar targets, which are point-sources, can be placed into the smaller payload point-source apertures without risk of significant vignetting (\S\ref{section-coalign}). Based on BCT specifications, we expect 1-sigma RMS pointing stability to be less than 7", while prior CU-LASP programs have traditionally out-performed this stability \cite{Egan23,Dolon25}. Pointing is controlled with a BCT XACT-50 and supported with a second co-boresighted star-tracker. 

\vspace{0.05in}
\noindent{\bf {\underline{The High-Voltage Power Supply:}}} 
\textcolor{black}{The MANTIS MCP detector is operated at a high voltage of $\approx$ -3125V. The HV power supply (HVPS) is an Ultravolt 6AA12-N4-M-I5 module driven by a custom LASP built HV control board. The control board drives the HVPS with options for both a fixed set of outputs and well as a DAQ-controlled variable output ranging from 0 -- -4000V. The variable output is the default state. The MANTIS HVPS design is derived from the SPRITE program\cite{Fleming19}}

\textcolor{black}{When not actively observing targets, the MANTIS detector is held in a STANDBY state of approximately -2500V - a level at which the detector is not active and is safe for transiting various non-operational states, such as over the poles or the South Atlantic Anomaly (SAA).}

\textcolor{black}{The SPRITE HVPS that the MANTIS HVPS is based off of was operated on a dummy resistive load for over three months in vacuum as part of flight qualification. A disassembly of the Ultravolt module does show voids in the manufacturer potting that, if aligned with a HV-carrying component and allowed to out-gas to coronal pressures, could result in arcing of the supply in space. On SPRITE, this was mitigated by over-potting the entire HVPS unit with Scotchweld 2216 epoxy, which had the added benefit of passivation of the silicone-based potting of the Ultravolt. For MANTIS we are investigating removing the silicone-based manufacturer potting and replacing it with a properly out-gassed, void-free, flight grade solution.} 

\section{The Performance of the MANTIS Science Instrument}\label{section-performance}

The MANTIS instrument is designed to observe the flux and flux variability of low-mass stars in the EUV through NUV/optical bandpasses. While the EUV channel is a spectrometer, only the total EUV irradiance is necessary for MANTIS to achieve its baseline science objectives. The EUV spectrograph is intentionally low resolution so as to limit the background equivalent flux (area of the detector subtended by the spectrum) of the EUV channel, and therefore maximize sensitivity. The expected EUV flux from all stars other than the Sun is small, and can only be estimated by models or the very limited data in the $EUVE$ archive (see above). The FUV and NUV/O flux is better known and the background limitations for those channels are not driving the sensitivity calculations. The driving requirements of the MANTIS science instrument channels are given in Table~\ref{tbl-requirements}.

\begin{table}[!ht]
\centering

\small 
\begin{tabularx}{\textwidth}{l p{2.1cm} X c}
\toprule
\textbf{Number} & \textbf{Description} & \textbf{Requirement} & \textbf{Current Estimate} \\
\midrule
INST 1 & EUV Sensitivity & The instrument shall have flux sensitivity ($> 5\sigma$) integrated over $150 - 250$~\AA\ of an average of $5\times10^{-15}$~c.g.s. in 100~ks & SNR = 43.2 \\
\addlinespace
INST 2 & FUV Sensitivity & The instrument shall have flux sensitivity ($> 5\sigma$) at 1250~\AA\ of $3\times10^{-15}$~c.g.s. per \AA\ in 100~ks & SNR = 13.8 \\
\addlinespace
INST 3 & NUV Sensitivity & The instrument shall have flux sensitivity ($> 5\sigma$) at 3500~\AA\ of $1\times10^{-13}$~c.g.s. per \AA\ in 100~ks (in sets of $\sim$300~s exposures) & SNR = 61.4 \\
\addlinespace
INST 4 & EUV Flare Sensitivity & The instrument shall have flux sensitivity ($> 3\sigma$) integrated over $150 - 250$~\AA\ of an average of $5\times10^{-14}$~c.g.s. per \AA\ in 300~s & SNR = 8.2 \\
\addlinespace
INST 5 & FUV Flare Sensitivity & The instrument shall have flux sensitivity ($> 3\sigma$) integrated over the $1250 - 1500$~\AA\ band of an average of $3\times10^{-14}$~c.g.s. per \AA\ in 120~s & SNR = 23.4 \\
\addlinespace
INST 6 & NUV Flare Sensitivity & The instrument shall have flux sensitivity ($> 3\sigma$) integrated over the $2200 - 3500$~\AA\ band of an average of $2\times10^{-14}$~c.g.s. per \AA\ in 120~s & SNR = 7.9 \\
\addlinespace
INST 7 & EUV Spectral Res. & The instrument shall have a filled-aperture spectral resolution in the EUV channel of less than 70~\AA\ to adequately separate geocoronal features & 37.1~\AA \\
\addlinespace
INST 8 & FUV Spectral Res. & The instrument shall have a filled-aperture spectral resolution in the FUV of $< 15$~\AA\ to isolate major emission lines & 2.2~\AA \\
\addlinespace
INST 9 & NUV Spectral Res. & The instrument shall have spectral resolution in the NUV channel of less than 200~\AA\ to fit an NUV spectral energy distribution & 15.5~\AA \\
\bottomrule
\end{tabularx}
 \caption[A summary of the driving MANTIS requirements] 
 { \label{tbl-requirements} 
{\em A summary of the driving performance requirements for the MANTIS instrument. As a SmallSat program, it is important for MANTIS to carry ample margin on these requirements to maintain a high risk posture.}}
\end{table}

This section first calculates the effective area and background rates from component analysis for each channel, and then verifies that the driving requirements in Table~\ref{tbl-requirements} are met by the MANTIS design. We start by first presenting the as-delivered performance of the EUV/FUV detector, then calculating the FUV and NUV/O channels effective area (A$_{eff}$) and sensitivities, then discussing the development of the EUV grating, and finally we calculate the EUV channel A$_{eff}$ and sensitivity. 

\begin{figure}[!t]
   \begin{center}
   \begin{tabular}{c}
 \includegraphics[width=0.99\textwidth,angle=0,trim={.5in 0.0in 00.0in 0.0in},clip]{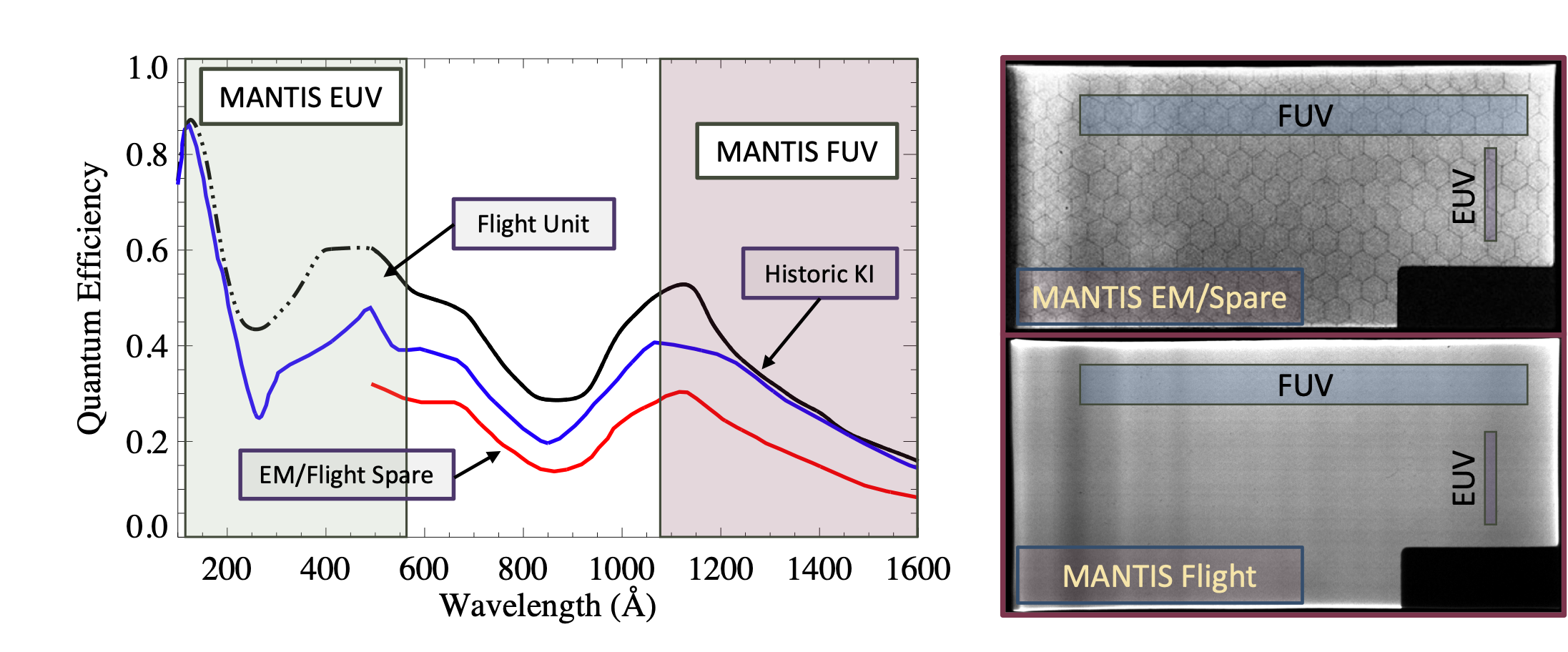} 
   \end{tabular}
 \end{center}
    \vspace{-0.15in}
   \caption[Quantum efficiency of the MANTIS MCP detectors]
  { \label{fig-detperf} 
{\em (Left) The quantum efficiency of the MANTIS EM/spare with borosilicate channel plates (red) slightly under-performs historical QEs for KI photocathodes (blue), however the flight detector represents the highest KI photocathode QEs ever recorded between 400 - 1100 \AA . The QE from 100 - 400 \AA\ could not be measured with the equipment currently operational at Sensor Sciences, therefore the dot-dashed lines are estimates. There is no QE enhancement at $\lambda$ $>$ 1200 \AA\ as the photon energy is less than half the KI bandgap. (Right) A deep flat-field of the EM/Spare (top) and flight MANTIS detectors. The EM/spare features the traditional hexagonal multi-fiber boundaries, while the new enhances channel plates show almost no image modulation. The dark rectangle in the lower right of each image is a region of the detector with no photocathode, where the zero-order of the EUV channel will land.}}
\vspace{-0.15in}
\end{figure}

\subsection{EUV/FUV Detector Performance}\label{section-detector}
The MANTIS EUV/FUV MCP detectors (flight and EM/spare) have both been completed and delivered to CU-LASP, where they are being stored under high vacuum ($<$ 10$^{-7}$ torr) pending testing with the instrument optics. The EM uses a borosilicate glass microchannel plate stack, which has exceptionally low background rates $\lesssim$ 0.05 events cm$^{-2}$ s$^{-1}$, while the flight unit uses a new form of leaded glass plates that have vastly reduced image modulation and improved quantum efficiency (D$_{QE}$) relative to traditional MCPs at the cost of slightly higher backgrounds $\sim$ 0.19 events cm$^{-2}$ s$^{-1}$. The delivered D$_{QE}$ for the flight unit is roughly 35\% higher than the EM (Figure~\ref{fig-detperf}). In both cases, instrumental and cosmic particle radiation is expected to dominate the intrinsic detector background, raising the on-orbit background rate to an estimated 2 events cm$^{-2}$ s$^{-1}$, or $\lesssim$ 0.4 events per 15 $\mu$m pixel per day. 

The implications of this new MCP plate technology is wide-reaching, as the enhanced D$_{QE}$ has been demonstrated to be consistently high for multiple photocathodes, greatly increasing instrument throughput with no increase in cost or complexity. Additionally, the reduced image modulation and fixed pattern noise typically caused by the multi-fiber bundle boundaries in MCPs reduces calibration uncertainties and signal-to-noise ceilings for high SNR observations (e.g. high-resolution stellar spectroscopy). MANTIS represents the first orbital flight program that these plates will be deployed on. 

Both MANTIS detectors have will operate with high voltage in the negative 3-3.2 kV range and have demonstrated resolution of $\sim$ 60 $\mu$m, which is significantly better than the 80 $\mu$m requirement. The maximum expected global count rate for any MANTIS target is less than 100,000 events per second, for which the dead time on the MANTIS detector is $\lesssim$ 5\%. The vast majority of targets will have count rates less than 3000 events per second, mostly from geocoronal sources. 

\begin{figure}[!t]
   \begin{center}
   \begin{tabular}{c}
 \includegraphics[width=0.99\textwidth,angle=0,trim={.0in 0.0in 00.0in 0.0in},clip]{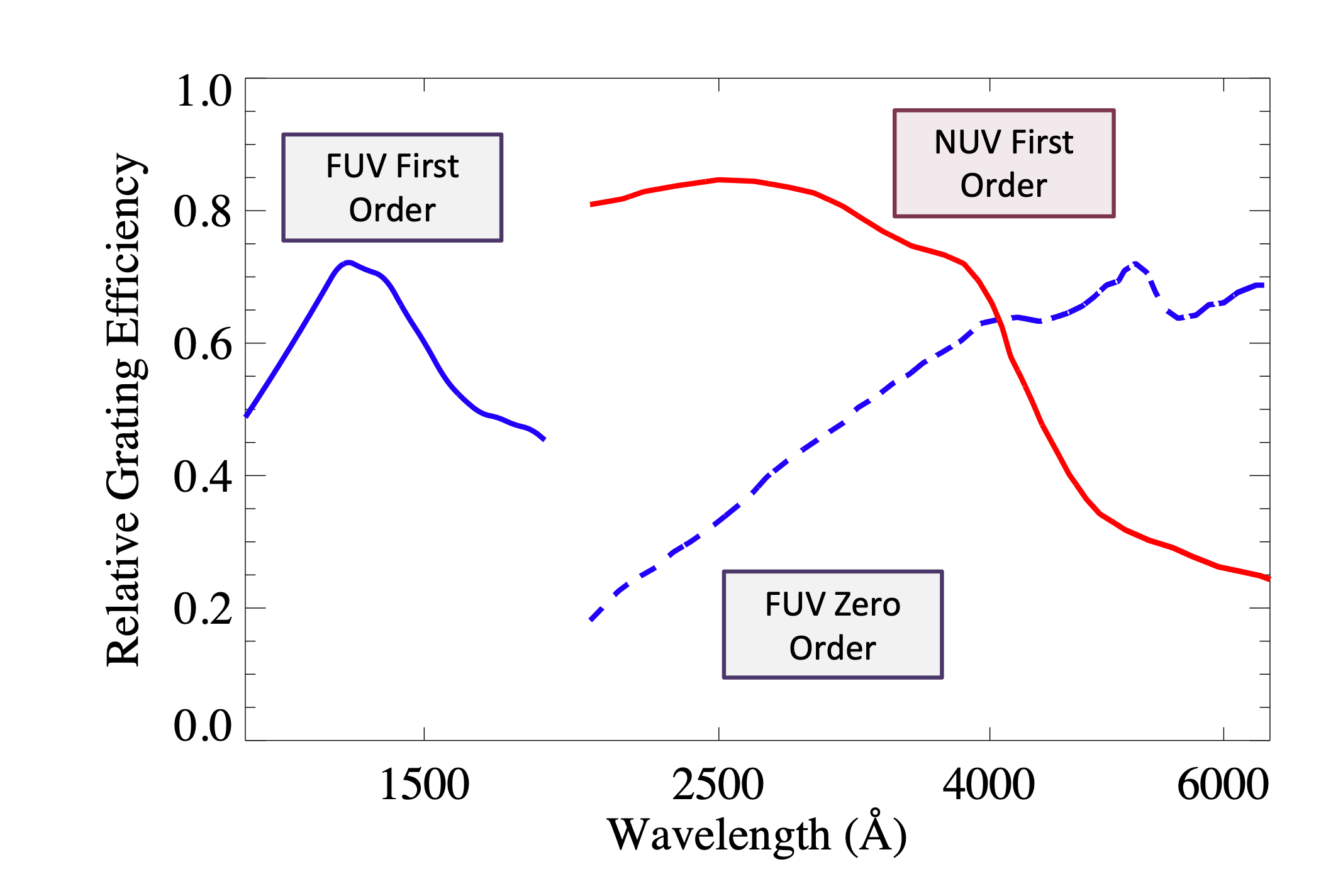} 
   \end{tabular}
 \end{center}
    \vspace{-0.15in}
   \caption[The diffraction efficiency of the MANTIS FUV/NUV grating]
  { \label{fig-fngrat} 
{\em The first diffracted order (blue solid) and zero-order (blue dashed) efficiency of the MANTIS/INFUSE FUV grating. The product of the NUV/O grating (red solid) and FUV zero-order feed into the NUV/O channel effective area calculation.}}
\vspace{-0.15in}
\end{figure}

\subsection{NUV/O and FUV Channel Effective Areas and Sensitivities}\label{section-fnuvperf}
\begin{figure}[!t]
   \begin{center}
   \begin{tabular}{c}
 \includegraphics[width=0.99\textwidth,angle=0,trim={.0in 0.0in 00.0in 0.0in},clip]{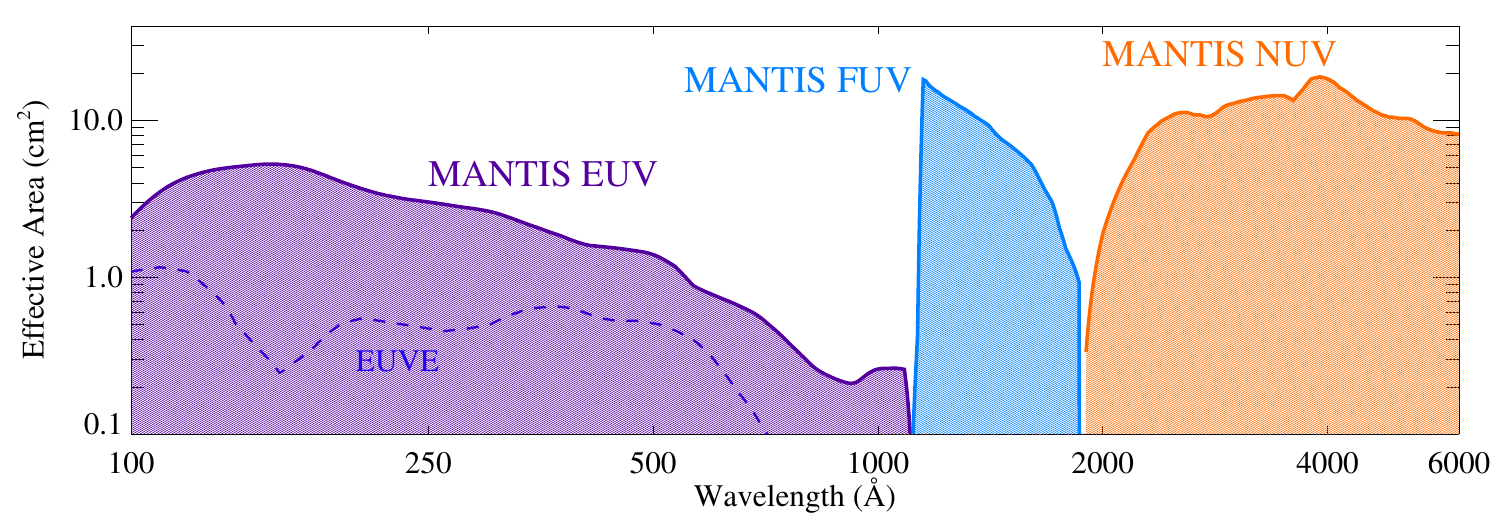} 
   \end{tabular}
 \end{center}
    \vspace{-0.15in}
   \caption[The effective areas of the MANTIS channels]
  { \label{fig-effas} 
{\em The effective areas of MANTIS for each channel, as derived from the component efficiencies. The $EUVE$ DS/S is included for comparison, though DS/S could not observe the entire bandpass shown in a single observation. MANTIS is the most efficient and sensitive spectrometer ever flown for the full bandpass of EUV astrophysics.}}
\vspace{-0.15in}
\end{figure}

The effective area (A$_{eff}$) of any spectroscopic channel is the product of the telescope collecting area (A$_{geo}$), the mirror coatings (R$_{m}$), the grating efficiency ($\epsilon_{g}$) and the detector quantum efficiency (D$_{QE}$). The D$_{QE}$ of the EUV/FUV detector is established in \S\ref{section-detector}, while the NUV/O detector D$_{QE}$ is taken from the CUTE program (which closely matched the CCD42-10 datasheet for the enhanced UV coating) \cite{Nell21}. The FUV/NUV/O grating efficiencies are taken directly from measurements by the INFUSE program, from which the MANTIS FUV/NUV/O grating is a replica \cite{Witt23,Haughton25JATIS}. The pre-coating efficiency curves of the delivered gratings have been measured and show initial values commensurate with the INFUSE gratings. The zero-order efficiency is applied to the NUV/O channel and compounded with the measured NUV/O efficiency of the Spectrum Scientific 300-250-012 300 groove mm$^{-1}$ blazed grating (Figure~\ref{fig-fngrat}). The mirror coatings for the MANTIS FUV and NUV/O channels are magnesium fluoride protected aluminum (MgF$_{2}$+Al) applied by Teledyne Acton Optics, with the coating curve taken from their catalog for the 1200 coating option (optimized for 1200 \AA ). A$_{geo}$ for the FUV/NUV/O channels is the collecting area of a 14 $\times$ 8 cm telescope with an 8\% estimated obscuration for the secondary mirror and mount structures, or 103.4 cm$^{2}$. The effective areas of all MANTIS channels are shown in Figure~\ref{fig-effas}, with the complexities of the calculations of the EUV channel presented in \S\ref{section-euvperf}. 

The sensitivities of the FUV and NUV/O channels are a function of the effective areas, background/noise rates, and exposure times, with the latter given in Table~\ref{tbl-requirements} for each requirement. In some cases, the requirements specify a single wavelength (e.g. INST 2 and INST 3) while in others it is over an integrated range. In each case, the area of the respective detector subtended by the integrated signal drives the background. The signal-to-noise ratios presented in Table~\ref{tbl-requirements} are calculated as:

\begin{equation}
    SNR=Signal/\sqrt{Signal + N_{Dark} + N_{Read}}
\end{equation}

where the read noise, N$_{read}$ is 0 for the EUV/FUV channels (owing to the use a photon-counting detector) and the product of the read noise multiplied by the number of exposures for the NUV/O channel.

As stated in \S\ref{section-fuvchannel}, the FUV channel dispersion is 21.4 \AA\ mm$^{-1}$, while the NUV/O is 155.1 \AA\ mm$^{-1}$ (\S\ref{section-nuvchannel}). A 250 \AA -wide bandpass in the FUV channel (INST 5), for instance, therefore covers an area of the detector approximately 11.7 mm wide by 0.1 mm high. At an expected in-flight background of 2 events cm$^{-2}$ s$^{-1}$ (\S\ref{section-detector}), this equates to 0.023 events s$^{-1}$ integrated across the 1250 - 1500 \AA\ bandpass. In a 100ks total observation (INST 1), this totals a background of $\sim$ 2300 counts of background. 

For the CCD this calculation is complicated by the read noise (estimated for now at 3.8 e$^{-}$ pixel$^{-1}$) and the requirement that long exposures be broken up into many smaller exposures to accommodate Earth occultations, avoid cosmic ray saturation, and to deal with target visibility. For Requirement INST 3, for instance, the extraction region is projected to be only $\sim$ 108 $\times$ 108 $\mu$m (8$\times$8 pixels). With a total dark rate of just 0.01 event pixel$^{-1}$ s$^{-1}$, this equates to a dark noise background of N$_{Dark}$ = 64,000 DN.  We assume that the maximum exposure time for the NUV/O CCD is 300 seconds to limit cosmic rays and to maximize temporal sampling for stellar flaring, however this will be refined as the concept of operations for MANTIS evolves. This amounts to $\approx$ 333 exposures for a 100ks total observations, for a read noise of n$_{Read}$ $\approx$ 81,000 DN. Total background over this 100ks observation is therefore $\sim$ 145,000 DN. 

The Signal in each case is the flux level set in the requirement multiplied by the integrated effective areas over the appropriate bandpasses. MANTIS carries large margins on all performance requirements, which is essential given the low-cost/high-risk nature of the program. This margin is available to trade throughout the program to maintain schedule and budget. 

\subsection{EUV Grating Performance}\label{section-euvgrating}
\begin{figure}[!t]
   \begin{center}
   \begin{tabular}{c}
 \includegraphics[width=0.99\textwidth,angle=0,trim={.0in 0.0in 00.0in 0.0in},clip]{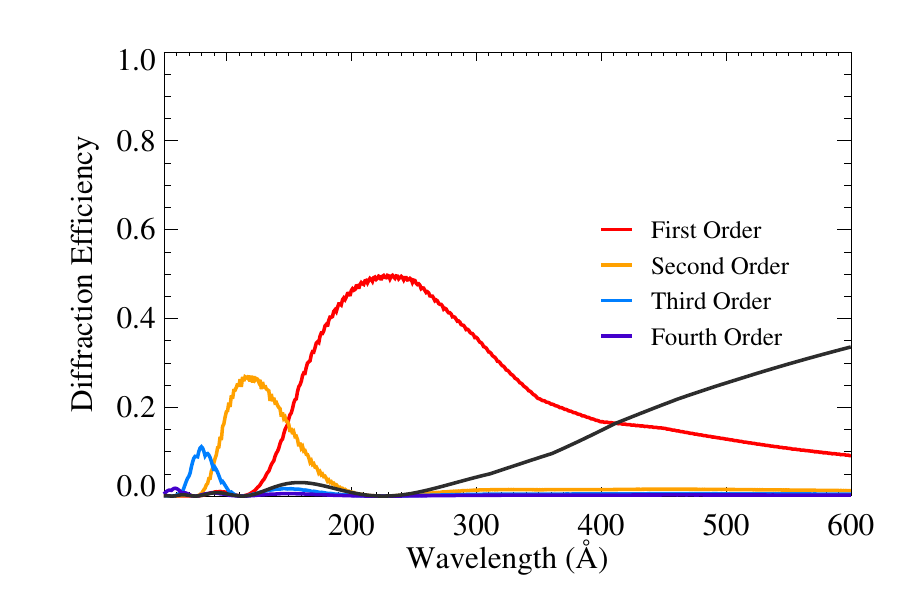} 
   \end{tabular}
 \end{center}
    \vspace{-0.15in}
   \caption[The rojected diffraction efficiency of the MANTIS EUV grating]
  { \label{fig-euvgratings} 
{\em The projected diffraction efficiency of the MANTIS EUV grating, including the effects of the Zr coating averaged over the power-weighted range of reflection angles. Third order (m=3) is almost entirely suppressed by the steep graze angles on the telescope, leaving primarily only two EUV spectral orders overlapping on MANTIS. The zero-order (m=0; black) curve is also shown.}}
\vspace{-0.15in}
\end{figure}

Three MANTIS EUV grating prototypes have been fabricated at PSU, with the best of the prototypes subsequently coated with 20 nm of zirconium (Zr). This test grating is currently at CU-LASP for testing. We chose a blaze angle of 0.9\adeg\ (225 \AA ) primarily to limit the total number of overlapping orders to $\sim$ 2 to reduce confusion (see Figure~\ref{fig-euvgratings}), while also avoiding having the m=2 peak order efficiency overlapping with the 304 \AA\ \ion{He}{2} airglow feature. 

The prototype grating being tested has a blaze angle of 1.4\adeg , which is higher than the 1.0\adeg\ flight-model target; a result of being ruled on crystalline silicon wafers that were readily available to the project. The impact of the larger blaze angle on the prototype optic is to shift the efficiency of the grating to longer wavelengths. This only reduces the total number of photons into the EUV channel by a small fraction ($<$ 10\%), as the MANTIS EUV spectrograph contains multiple overlapping spectral orders, and a shift of the first-order blaze wavelength correspondingly shifts the m=2,3, etc orders as well.  The flight grating substrates required a custom cut to the appropriate $<$111$>$ crystal plane orientation. These are expected to be delivered to PSU in the summer of 2026, with the flight grating test rulings occurring as of the publication of this paper. The prototype grating testing has been complicated by challenges related to the CU-LASP Manson soft X-ray light source, however the longer wavelength end of the grating efficiency curve has been measured and implies that the delivered gratings will exceed expectations (Figure~\ref{fig-euvgratings}). Full results will be published at a later date. 

\subsection{EUV Channel Effective Area}\label{section-euvperf}
\begin{figure}[!t]
   \begin{center}
   \begin{tabular}{c}
 \includegraphics[width=0.99\textwidth,angle=0,trim={.0in 0.0in 00.0in 0.0in},clip]{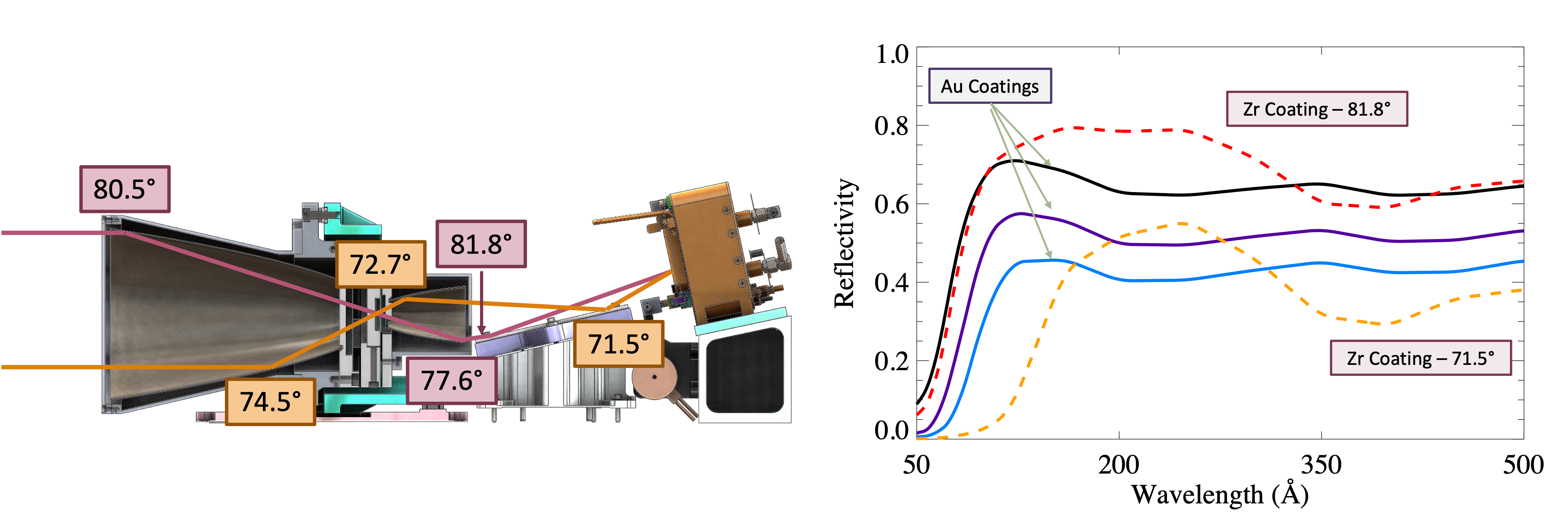} 
   \end{tabular}
 \end{center}
    \vspace{-0.15in}
   \caption[A summary of the reflectance estimates for the EUV channel optics]
  { \label{fig-grazeref} 
{\em (Left) a CAD cartoon showing a grazing incidence beam at the outer edge of the top of the primary mirror (pink) relative to the inner illuminated edge of the lower half of the primary mirror (orange). The larger the angle of incidence, the higher the reflectivity. (Right) The reflectance as a function of EUV wavelength of the gold (Au) coated telescope mirrors (72.7\adeg\ -- 80.5\adeg ; solid) and zirconium (Zr) coated grating (71.5\adeg\ -- 81.8\adeg ; dashed)}}
\vspace{-0.15in}
\end{figure}

The EUV channel A$_{eff}$ is complicated by the highly variable incident angles, and therefore reflectivity, of the EUV photons off of grazing incidence optics. The reflectance as a function of angle for various wavelengths is presented in Figure~\ref{fig-grazeref}. At the leading edge of the HB primary mirror, for instance, the incidence angle for on-axis light is 80.5\adeg, while at the inner edge of the clear aperture of the primary mirror this falls to just 74.5\adeg. Incidence angles on the secondary mirror range from 72.7 to 77.6\adeg\, while the grating spans 71.5 -- 81.8\adeg. This leads to significant variation in throughput depending on the location of each incident photon. \textcolor{black}{Polarization effects from high angle-of-incidence reflections can also be important, though at the MANTIS angles-of-incidence these are only $\sim$ 10\% effects. As the sources are not expected to be strongly polarized, nor is polarization important to the MANTIS science, this splitting can be ignored. All reflectances are assumed to be the average "unpolarized" values.}

To calculate the EUV A$_{eff}$, we integrate the throughput over all possible incident angles into the clear aperture of the telescope. From this analysis, the top half of the telescope aperture carries over 65\% of the total EUV-channel power (due to the incident angle of light through that part of the telescope onto the grating), with the outer edges of the primary mirror having more impact on the A$_{eff}$ than the inner region. The innovative telescope design, advanced etched silicon diffraction grating, and KI photocathode advanced MCP detector all combine to make MANTIS the most efficient, and most sensitive EUV spectrograph to-date for astrophysics. 

\subsection{Co-alignment of the MANTIS Channels}\label{section-coalign}
MANTIS will observe low mass stars over long baselines to measure both the full UV irradiance from 100 - 6400 \AA\ as well as the flare properties. Target stars will be observed while MANTIS is in the shadow of the Earth, where the geocoronal \lya, \ion{He}{1} and \ion{He}{2} emission is at a minimum. Long baseline observations $\sim$ 100ks will therefore necessarily be split over several weeks, as only $\approx$ 25ks of shadow time is available per calendar day, after accounting for efficiency losses due to the SAA, downlinks, etc. MANTIS is capable of daytime observations as well, provided the solar array can be effectively oriented towards the sun. Such EUV dayside observations may require additional calibration to deal with increased geocoronal emissions and potentially scattered light. The FUV and NUV/optical channels are much less affected by dayside airglow, with the primary impact being increased noise (factor of 3$\times$ compared to baseline estimates) from 1150~--~1250~\AA.  

\begin{figure}[!t]
   \begin{center}
   \begin{tabular}{c}
 \includegraphics[width=0.99\textwidth,angle=0,trim={.0in 0.0in 00.0in 0.0in},clip]{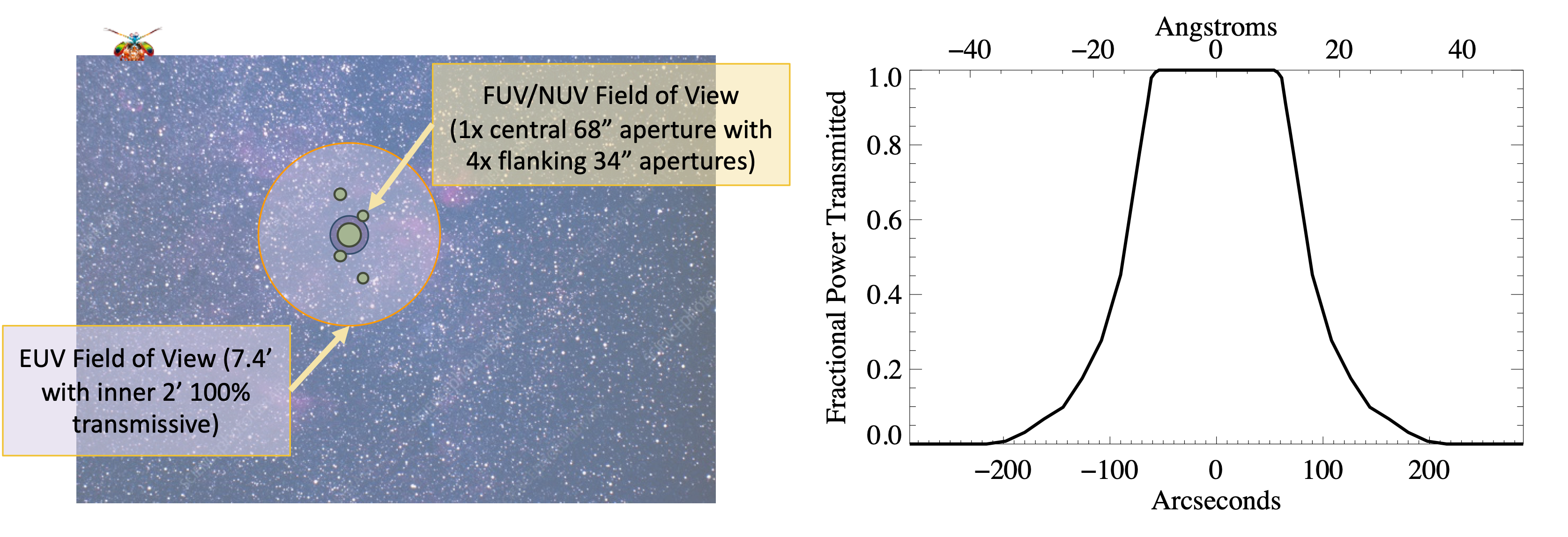} 
   \end{tabular}
 \end{center}
    \vspace{-0.15in}
   \caption[A depiction of the channel co-alignment and EUV aperture function]
  { \label{fig-transmission} 
{\em (Left) A cartoon of the EUV FoV superimposed on the sky with the central 2' region in darker purple. The FUV/NUV/O aperture mask if overlaid in green with the 68" central aperture centered on the middle of the EUV aperture. (Right) The modeled transmission function for the EUV PFA. The top axis shows the transmission function in angstroms, showing how the EUV geocoronal emission lines will spread over a portion of the EUV bandpass.}}
\vspace{-0.15in}
\end{figure}

The most important driver of the calibration performance  of MANTIS is to achieve and then maintain co-alignment between all of the spectral channels. This is especially important not only for increasing observational efficiency, but also essential for ensuring that active stars are monitored simultaneously across the entire MANTIS waveband to record panchromatic flare energies. The co-alignment requirement is driven by the EUV-channel pinhole aperture transmission function, which has a 2\amin\ unvignetted region within a 7.6\amin\ total field-of-view (Figure~\ref{fig-transmission}). As long as any of the FUV/NUV/O apertures can be effectively steered into this central region and maintained there, the two channels are adequately aligned. 

The alignment procedure will be carried out using a 10-inch diameter collimated beam that effectively illuminates both telescope apertures simultaneously. The FUV/NUV/O telescope will be installed first with only mechanical tolerances establishing the position. The EUV telescope will then be installed with no grating or EUV/FUV detector, but instead a laboratory CMOS camera near each telescope focus. The collimated point-source is then steered into the central aperture of the FUV/NUV/O PSA (Figure~\ref{fig-teleslit}) and scanned in pitch and yaw to ensure it is placed at the center of the 68\asec\ aperture. The beam is then steered within precision micrometers into the EUV telescope PFA, and again scanned in pitch and yaw to find the center of the PFA. Using the displacements of the beam steering micrometers, the angular offset between the two telescopes can be calculated, and the EUV telescope appropriately shimmed to converge on alignment. This procedure is repeated until the shims bring the two telescopes to within 10\asec , with the remaining alignment margin held against vibration/launch/thermal shifts. 

Following co-alignment, the instrument will be installed into the CU-LASP Long Tank facility - a vacuum collimator with a ~ 45 cm diameter beam fed by a UV hollow cathode light source. This source and the gold-coated, normal-incidence Long Tank optics will produce only a minimal amount of EUV light, however spectral features of Argon and Neon at the edge of the MANTIS EUV bandpass have been detected with the prior CU-LASP sounding rocket, DEUCE \cite{Erickson19}. The Long Tank will provide a baseline pre-vibe spectrum in each channel (including zero-order in the EUV channel) with a simulated point-source. From there, the system will be placed in a second vacuum chamber and directly illuminated with a hollow cathode light source, simulating diffuse emission with $\geq$ 100$\times$ higher incident intensity without the collimating optics. This will aid in stray light testing, but also provide filled-aperture spectroscopy in each channel simultaneously. These measurements are repeated post-vibe and at other testing stages in the MANTIS development. For EUV light to reach the detector, the door on the MCP housing must be deployed, however the MgF$_{2}$ window will allow the FUV light into the system, while the NUV/O has no aperture cover.

\section{Conclusions}\label{section-conclusion}
The MANTIS SmallSat represents the first EUV-sensitive orbital astrophysics mission in nearly 30 years \textcolor{black}{(following $EUVE$ and CHIPSat\cite{CHIPsat})}, and is projected to have $\sim$ 5$\times$ the throughput as the $EUVE$-DS/S. With nearly contiguous spectral coverage from 100 -- 6400 \AA , MANTIS will sample the full UV irradiance and activity of a diverse sample of stars to evaluate the potential of such stars to support habitable exoplanets. The MANTIS optical design leverages heritage from prior CU-LASP UV CubeSats SPRITE and CUTE, but also serves as a technology development platform for a new type of grazing incidence telescope, advanced UV diffraction grating fabrication techniques, and microchannel plate detectors. All of these technologies were originally studied for the ESCAPE Small Explorer proposal submitted in 2019, which was awarded a Phase A concept study by NASA and will be proposed again in 2026 \cite{France20, Fleming21}. The MANTIS grating technology is also baselined for an upcoming sounding rocket program, MOBIUS \cite{Haughton25JATIS}, and has the potential to be leveraged for the Habitable Worlds Observatory. The MANTIS SmallSat program is serving as a powerful development and demonstration platform for these enabling technologies. The ESPA-sized spacecraft is also a first for a NASA-funded APRA program, demonstrating that at these scales and the MANTIS risk profile, cost is not driven by volume or mass. Instead, the MANTIS size represents a risk-reduction, alleviating the tight constraints of a conventional CubeSat volume. 

The MANTIS program is currently entering the first stages of integration and testing, with an anticipated launch readiness \textcolor{black}{date} in 2027. 

\subsection*{Disclosures}
The authors declare that there are no financial interests, commercial affiliations, or other potential conflicts of interest that could have influenced the objectivity of this research or the writing of this paper.

\subsection* {Code, Data, and Materials Availability} 
No custom analysis code used in the creation of this paper is publicly available. Signal-to-noise and other calculations were carried out following established methodologies, while plots are customized to this work as a means to illustrate a concept. Any questions or clarifications regarding these calculations may be referred to the corresponding author.

\subsection* {Acknowledgments}
This paper includes optical designs, technology performance, and lessons learned derived from experiences on suborbital instrument, technology development, and Roman Technology Fellowship grants from the National Aeronautics and Space Administration (NASA), award No(s) 80NSSC19K0450, 80NSSC19K0345, 80NSSC22K1521, and 80NSSC24K0231, with the primary MANTIS award 80NSSC24K0304.

\clearpage

\bibliography{mybib} 





\end{spacing}
\end{document}